\documentclass[a4paper,10pt]{article}

\usepackage{hyperref}
\usepackage[utf8]{inputenc}
\usepackage{amssymb}
\usepackage{amsmath}
\usepackage{graphicx}
\usepackage{xcolor}
\usepackage{natbib}

\usepackage{algorithm}
\usepackage{algpseudocode}
\usepackage{algorithmicx}

\usepackage[left=2cm,top=2.5cm,right=2cm,bottom=2.5cm]{geometry}

\hypersetup{colorlinks,linkcolor= {blue},citecolor= {blue}}

\title{A comparative study of data- and image- domain LSRTM under velocity-impedance parametrization}

\author{Pengliang Yang$^\dagger$ and Zhengyu Ji\\
  School of Mathematics, Harbin Institute of Technology, 150001, Harbin, China\\
  $^\dagger$E-mail: ypl.2100@gmail.com
}

\begin{document}

\maketitle

\begin{abstract}
  Least-squares reverse time migration (LSRTM) is one of the classic seismic imaging methods to reconstruct model perturbations within a known reference medium. It can be computed in either data or image domain using different methods by solving a linear inverse problem, whereas a careful comparison analysis of them is lacking in the literature. In this article, we present a comparative study for multiparameter LSRTM in data- and image- domain in the framework of SMIwiz open software.   Different from conventional LSRTM for recovering only velocity perturbation with variable density, we focus on simultaneous reconstruction of velocity and impedance perturbations after logorithmic scaling,  using the first-order velocity-pressure formulation of acoustic wave equation. 
  The first 3D data-domain LSRTM example has been performed to validate our implementation, involving expensive repetition of Born modelling and migration over a number of iterations. As a more cost-effective alternative, the image-domain LSRTM is implemented using point spread function (PSF) and nonstationary deblurring filter. Dramatic disctinctions between data and image domain methods are discovered with 2D Marmousi test: (1) The data-domain multiparameter inversion provides much better reconstruction of reflectivity images than image-domain approaches, thanks to the complete use of Hessian in Krylov space; (2) The poor multiparameter image-domain inversion highlights the limitation of incomplete Hessian sampling and strong parameter crosstalks, making it difficult to work in practice; (3) In contrast, monoparameter image-domain inversion for seismic impedance is found to work well. These observations have been further validated on Viking Graben Line 12 dataset.
\end{abstract}

\section{Introduction}
Seismic imaging techniques like reverse time migration (RTM) \citep{Baysal_1983_RTM, McMechan_1983_MET, Whitmore_1983_IDM} and full waveform inversion (FWI) \citep{Virieux_2009_OFW} utilize wave modeling to probe the Earth's complex subsurface for geophysical exploration. The estimation of subsurface parameters is fundamentally a least-squares inverse problem \citep{Tarantola_1982_IPQ}. Seminal work by \citet{Lailly_1983_SIP} and \citet{Tarantola_1984_ISR} under the acoustic approximation revealed that calculating the FWI gradient for different model parameters strongly resembles a seismic migration process \citep{chavent1999optimal}. When a sufficiently accurate background velocity model is available, this framework can also estimate model perturbations, motivating the development of linearized waveform inversion to derive seismic images from reflection data \citep{Tarantola_1984_LIS}. By iteratively minimizing the misfit between observed and simulated data, this method effectively suppresses migration artifacts and compensates for uneven illumination caused by finite acquisition apertures, thereby retrieving a quantitative reflectivity model beyond a simple structural image \citep{Symes_2008_ALI}.

Linearized waveform inversion was subsequently extended to multiparameter acoustic and elastic media \citep{LeBras_1985_PhD, lebras1988iterative, Beylkin_1990_LIS} and later to viscoacoustic settings \citep{Blanch_1995_efficient}. Early literature often referred to this approach as true amplitude migration \citep{Black_1993_TAI, Rousseau_2000_TAM, Jin_2002_MPT, Zhang_2009_PIR}. Today, it is commonly termed data-domain least-squares reverse time migration (LSRTM), a nomenclature that reflects its evolution from least-squares Kirchhoff migration \citep{schuster1993least, chavent1999optimal, Nemeth_1999_LSQ} to its generalization with two-way wave equations \citep{Plessix_2004_FDF, dai2011least, Zhang_2015_LSM}. Elastic LSRTM has recently gained significant traction \citep{chen2017elastic, rocha2018elastic, yang2020elastic}, with numerous studies emphasizing the critical role of density variations in both acoustic \citep{Yang_2016_lsrtm, Lu_2018_approximate, Farias_2022_vector} and elastic inversions \citep{Qu_2018_ELSRTM, Sun_2018_ELSRTM, Chen_2019_LSRTM}, including extensions to other domains \citep{Farshad_2020_extended}. Various preconditioning strategies have also been proposed to accelerate its convergence \citep{chen2017elastic, Farshad_2021_accelerating}.

Classical LSRTM in data domain normally requires tens of iterations of the Born modelling and migration, which poses a significant computational challenge to process 3D field data. The image-domain LSRTM, sometimes referred to as migration deconvolution \citep{hu2001poststack,fletcher2016least}, reformulates the normal equation of data-domain problem as a new objective function. Image-domain approaches demand the accessibility to the Hessian matrix, which may be partially sampled using point spread functions (PSFs) \citep{lecomte2008resolution,fletcher2016least}. In recent years, the method has received intensive attention due to practical values in seismic imaging \citep{osorio2021migration,yang2022efficient,zhang2023tutorial,Sheng_2023_Image}. In retrieving reflectivity images, the deblurring filtering approach stands out due to significance of the computational efficiency \citep{guitton2004amplitude,aoki2009fast,dai2011least}, using effective filter to mimic the impact of inverse Hessian.

Despite the maturity of image-domain LSRTM, its application to multiparameter inversion remains largely unexplored. Furthermore, while most LSRTM studies focus on velocity inversion, often with variable density \citep{Farias_2022_vector}, a direct comparison between data- and image-domain methods for impedance parameter inversion is notably absent.

In this work, we perform a comparative analysis of linearized waveform inversion using a velocity-impedance parameterization. To account for the significant amplitude differences between parameters, we incorporate logarithmic scaling within the linear inversion. A suite of LSRTM implementations—encompassing both data-domain and image-domain approaches—has been developed within the SMIwiz open-source software framework \citep{Yang_2024_SMIwiz}, based on a first-order vector acoustic wave equation. We validate our implementation using 2D Marmousi and 3D Overthrust benchmarks, highlighting the distinct characteristics of each method. Additionally, we provide a mathematical analysis using the $2\times 2$ block matrix inverse formula to elucidate the fundamental distinctions between multiparameter and single-parameter inversion.

\section{Scattering equation under $\ln V_p-\ln I_p$ parametrization}

The acoustic wave equation is a linear partial differential equation of the form
\begin{equation}
  \underbrace{\begin{bmatrix}
      \rho\partial_t & \nabla\\
      \kappa\nabla\cdot & \partial_t
  \end{bmatrix}}_{A(m)} \underbrace{\begin{bmatrix}
      \mathbf{v}\\
      p
  \end{bmatrix}}_{u}=\underbrace{\begin{bmatrix}
      f_\mathbf{v}\\
      f_p
  \end{bmatrix}}_f,
\end{equation}
where $A(m)$ is a linear wave operator depending on the model parameter $m$ which is a function of medium properties (density $\rho$,  P wave speed $V_p$, impedance $I_p=\rho V_p$ and bulk modulus $\kappa=\rho V_p^2$). The vectors $u=[\mathbf{v},p]^\mathrm{H}$ and $f=[f_\mathbf{v},f_p]^\mathrm{H}$ collect the wavefields and the sources for particle velocities and pressure, respectively.
Perturbing the reference model $m_0$ with $\delta m$ leads to the scattering field $\delta u$ besides the background field $u_0$ ($A(m_0) u_0=f$) satisfying
\begin{equation}\label{eq:scattereq}
  A(m_0)\delta u = -\delta m\cdot \partial_m A(m_0) u_0,
\end{equation}
which is obtained using first order Born approximation while dropping higher order terms for the perturbed wave equation $A(m_0+\delta m)(u+\delta u)=f$.
Equation~\eqref{eq:scattereq} reveals that the scattering field $\delta u$  has a linear dependence on the model perturbation $\delta m$.

We apply a logarithmic transformation of all physical parameters as the model parameters  such that
\begin{equation}\label{eq:ln}
  \frac{\partial A(m)}{\partial \ln\rho}=\frac{\partial A(m)}{\partial \rho}\frac{\partial\rho}{\partial\ln\rho}=\rho\begin{bmatrix}
  \partial_t & 0\\
  0 & 0
  \end{bmatrix}, \quad
  \frac{\partial A(m)}{\partial \ln\kappa}=\frac{\partial A(m)}{\partial \kappa}\frac{\partial\kappa}{\partial\ln\kappa}=\kappa\begin{bmatrix}
  0 & 0\\
  \nabla\cdot & 0
  \end{bmatrix}.
\end{equation}
The logarithmic transformation of density $\rho$ and bulk modulus $\kappa$ are related to  wavespeed and impedance through
\begin{equation}
  \begin{cases}
    \ln \rho=\ln I_p - \ln V_p\\
    \ln \kappa=\ln I_p+\ln V_p
  \end{cases}.
\end{equation}
where the multiplication between various physical parameters is converted to be addition of their log-scaled counterparts, therefore very convenient for efficient computation. The chain rule leads to
\begin{subequations}\label{eq:chainrule2}
  \begin{align}
    \frac{\partial A(m)}{\partial \ln I_p}
    =&\frac{\partial A(m)}{\partial \ln\rho}\frac{\partial \ln\rho}{\partial \ln I_p}+\frac{\partial A(m)}{\partial \ln\kappa}\frac{\partial\ln\kappa}{\partial\ln I_p}
    =\frac{\partial A(m)}{\partial \ln\rho}+\frac{\partial A(m)}{\partial \ln\kappa},\\
    \frac{\partial A(m)}{\partial\ln V_p}
    =&\frac{\partial A(m)}{\partial \ln\rho}\frac{\partial \ln\rho}{\partial \ln V_p}+\frac{\partial A(m)}{\partial \ln\kappa}\frac{\partial\ln\kappa}{\partial\ln V_p}
    =-\frac{\partial A(m)}{\partial\ln\rho} + \frac{\partial A(m)}{\partial\ln \kappa}.
  \end{align}  
\end{subequations}
The model parametrization by $(m_1,m_2)=(\ln V_p,\ln I_p)$ yields
\begin{equation}
  \delta m\cdot \partial_m A(m) u_0
  =\delta m_1\frac{\partial A(m)}{\partial \ln V_p}u_0
  + \delta m_2\frac{\partial A(m)}{\partial \ln I_p}u_0
  =\begin{bmatrix}
  (\delta m_2-\delta m_1)\rho\partial_t\mathbf{v}\\
  (\delta m_2+\delta m_1)\kappa \nabla\cdot \mathbf{v}
  \end{bmatrix},
\end{equation}
so that  the scattering equation \eqref{eq:scattereq} reads explicitly
\begin{equation}\label{eq:scattereqln}
  \underbrace{\begin{bmatrix}
      \rho\partial_t & \nabla\\
      \kappa\nabla\cdot & \partial_t
  \end{bmatrix}}_{A(m_0)} \underbrace{\begin{bmatrix}
      \delta \mathbf{v}\\
      \delta p
  \end{bmatrix}}_{\delta u}=\begin{bmatrix}
  -(\delta m_2-\delta m_1)\rho\partial_t \mathbf{v}\\
  -(\delta m_2+\delta m_1)\kappa \nabla\cdot \mathbf{v}
  \end{bmatrix},
\end{equation}
where we denote $u_0=[\mathbf{v}, p]^\mathrm{H}$ and $\delta u=[\delta \mathbf{v}, \delta p]^\mathrm{H}$.

\section{LSRTM in data and image domain}

Data-domain LSRTM, also referred to linearized waveform inversion, is to retrieve the model perturbation $\delta m$, assuming a reference model $m_0$ is known. It minimizes the misfit between the reflections $\delta d$ and the Born modelled data 
\begin{equation}
  J(\delta m) = \frac{1}{2}\|L[m_0]\delta m - \delta d\|^2,
\end{equation}
where the synthetic data recorded by receivers  is defined via $L[m_0]\delta m := R\delta u $ through a restriction operator $R$, which projects the wavefield within the reference medium $m_0$ to receiver positions. 
The gradient of the misfit functional with respect to $\delta m$ is
\begin{equation}\label{eq:LtL}
  \frac{\partial J(\delta m)}{\partial \delta m}=L^\mathrm{H} [m_0] (L[m_0]\delta m-\delta d),
\end{equation}
where $L^\mathrm{H}[m_0]$ is the adjoint of Born modelling.
For this reason, Born modelling operator $L[m_0]$ is sometimes referred to as reverse time demigration \citep{Symes_2008_ALI,Zhang_2015_LSM}.
The optimality condition $\partial J(\delta m)/\partial\delta m=0$ yields
\begin{equation}\label{eq:normal}
  \underbrace{L^\mathrm{H}[m_0] L[m_0]}_H\delta m= \underbrace{L^\mathrm{H}[m_0] \delta d}_{m_{rtm}}.
\end{equation}
The right hand side of the above normal equation is nothing else than the RTM image $m_{rtm}=L^\mathrm{H}[m_0] \delta d$, obtained by
 mapping the reflection data $\delta d$ back to the image domain. 
Equation~\eqref{eq:gradlsm} reveals that the image $m_{rtm}$ is parametrization dependent. Thus, it is a generalization of conventional RTM image computed via zero-lag cross-correlation imaging condition.
Equation \eqref{eq:normal} is usually solved using conjugate gradient (CG) methods to invert the symmetric positive definite (SPD) Hessian matrix $H=L^\mathrm{H}[m_0]L[m_0]$.

The expression in \eqref{eq:LtL} is abstract and formal, showing the gradient of the misfit is equivalent to the application of $L^\mathrm{H}[m_0]$ to $L[m_0]\delta m-\delta d$. On the other hand, Appendix \ref{sec:mig} reveals that the gradient of the misfit can be efficiently constructed via cross-correlation between the forward background field and an adjoint field, while $\delta d-L[m_0]\delta m$ is the right hand side of the adjoint equation \eqref{eq:adjeq}.
In other words, the action of $L^\mathrm{H}[m_0]$ to an input vector will be built in a matrix free manner via cross-correlation, feeding the negative input vector as the adjoint source. Under the model parametrization $(m_1,m_2)=(\ln V_p,\ln I_p)$, the relation in \eqref{eq:chainrule2} leads to
\begin{subequations}
  \begin{align}
    \frac{\partial J(\delta m)}{\partial \delta\ln I_p}
    =&\frac{\partial J(\delta m)}{\partial \delta\ln\rho}+\frac{\partial J(\delta m)}{\partial \delta \ln\kappa},\\
    \frac{\partial J(\delta m)}{\partial\delta\ln V_p}
    =&-\frac{\partial J(\delta m)}{\partial\delta\ln\rho} + \frac{\partial J(\delta m)}{\partial \delta\ln \kappa}.
  \end{align}
\end{subequations}
where  $\partial J(\delta m)/\partial\delta\ln \rho$ and  $\partial J(\delta m)/\partial\delta\ln \kappa$ can be computed using the explicit expressions in equation~\eqref{eq:gradrho} and \eqref{eq:gradkappa}.

Data-domain LSRTM requires repetitive migration and demigration in every iteration, which can be computationally expensive.
As a cost effective alternative, image-domain LSRTM reformulates the normal equation \eqref{eq:normal} into a new inverse problem
\begin{equation}\label{eq:migdeconobj}
  \min_{\delta m}\|H\delta m - m_{rtm}\|^2,
\end{equation}
assuming the accessibility of Hessian. For a model of $N$ unknowns, the size of Hessian matrix $H$ is $N^2$, which is prohibitively large to be stored for algebraic manipulations.
This implies that a practical image-domain LSRTM can only supply an approximate solution to the data-domain approach, by parsimonious storage of the Hessian or making approximations in computation under certain assumptions. These involves sampling and interpolating Hessin using the concept of point spread function (PSF) \citep{lecomte2008resolution,fletcher2016least}, as well as effectively mimicking the impact of inverse Hessian through the so-called deblurring filter.

To solve the above linear inverse problems, the CGNR (Conjugate Gradient Normal Residual) method has been employed due to its numerical robustness and efficiency. The convergence rate of CGNR may be further improved, combining a pseudo-Hessian preconditioner (Appendix \ref{sec:pseudohessian}), leading to the preconditioned CGNR method  \citep[Algorithm 9.7]{Saad_2003_IMS}.  The image-domain inversion can be started with the Hessian information built using Born modelling and migration. The PSFs can then be obtained by applying Hessian to an image of only a number of evenly distributed scatter points. When the input image is $m_{rtm}$, we obtain a remigrated image with the effect of Hessian. These results can then be used to perform image-domain inversion, see Appendix~\ref{sec:psf} and \ref{sec:fft}.

To analyze the efficiency and quality of these methods, we implemented both data-domain and image-domain LSRTM within the SMIwiz open-source software framework \citep{Yang_2024_SMIwiz}. SMIwiz utilizes a reverse communication interface for efficient multi-dimensional seismic modeling and imaging, maintaining a concise, modular structure that is particularly effective for the matrix-vector products in linearized inversion. The software incorporates routines from Seismic Unix for parameter parsing and employs a bijective shot-to-CPU mapping, enabling parallel execution with near-perfect performance scaling, as demonstrated in previous studies on RTM and nonlinear FWI \citep{Yang_2024_SMIwiz, Ji_2025_SMIwiz2}.

This work presents the first complete implementation of linearized inversion in SMIwiz for reflectivity imaging. In the data-domain, all shots are processed independently during modeling; the results are then reduced across all processors to form the descent direction in a matrix-free fashion. For the image-domain approach, computation switches to a single processor once the generalized RTM image is constructed. For further details on the software architecture, see \citet{Yang_2024_SMIwiz} and \citet{Ji_2025_SMIwiz2}.

\section{Numerical validation}

\subsection{2D Marmousi model}

\begin{figure}[htbp]
  \centering
  \includegraphics[width=\linewidth]{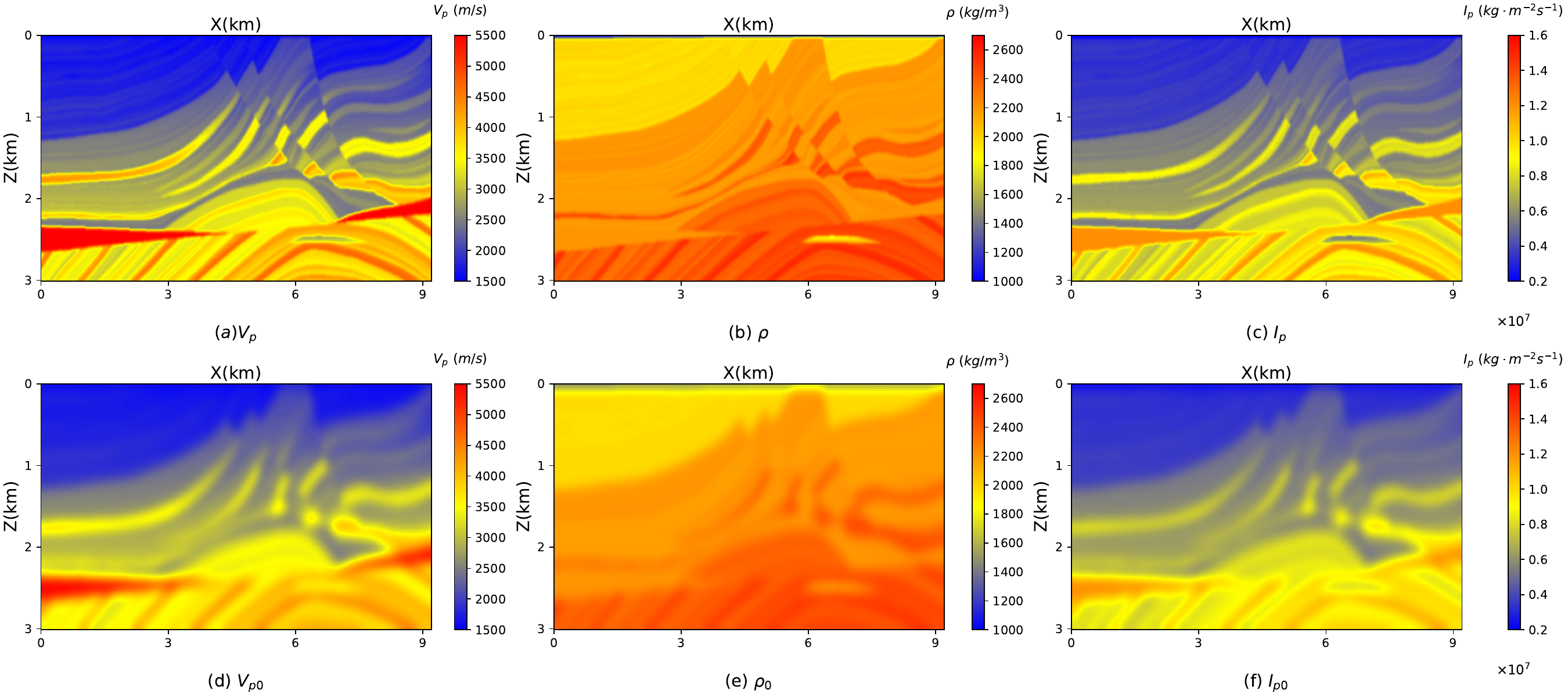}
  \caption{The true Marmousi model (1st row), smoothed one (2nd row) associated with velocity (1st column), density (2nd column) and impedance (3rd column)}\label{fig:marmousi}
\end{figure}

We first validate our implementation using the 2D Marmousi model with variable velocity and density, shown in Figure~\ref{fig:marmousi}a and ~\ref{fig:marmousi}b. The model dimensions are $n_x\times n_z=767\times 251$ with a grid spacing of $\Delta x=\Delta z=12$. Linearized inversion begins with smoothed versions of the true models (Figure~\ref{fig:marmousi}d and ~\ref{fig:marmousi}e). As our data-domain LSRTM employs $V_p-I_p$  parameterization rather than the provided $V_p$ and $\rho$, the starting impedance model (Figure~\ref{fig:marmousi}f) is derived by multiplying the smoothed $V_p$ and $\rho$ models (Figure~\ref{fig:marmousi}c).
The acquisition geometry consists of 55 shots evenly distributed at a depth of 5 m. Each shot is recorded by 1801 receivers at a depth of 10 m with a 5 m spacing.

\begin{figure}[htbp]
  \centering
  \includegraphics[width=0.8\linewidth]{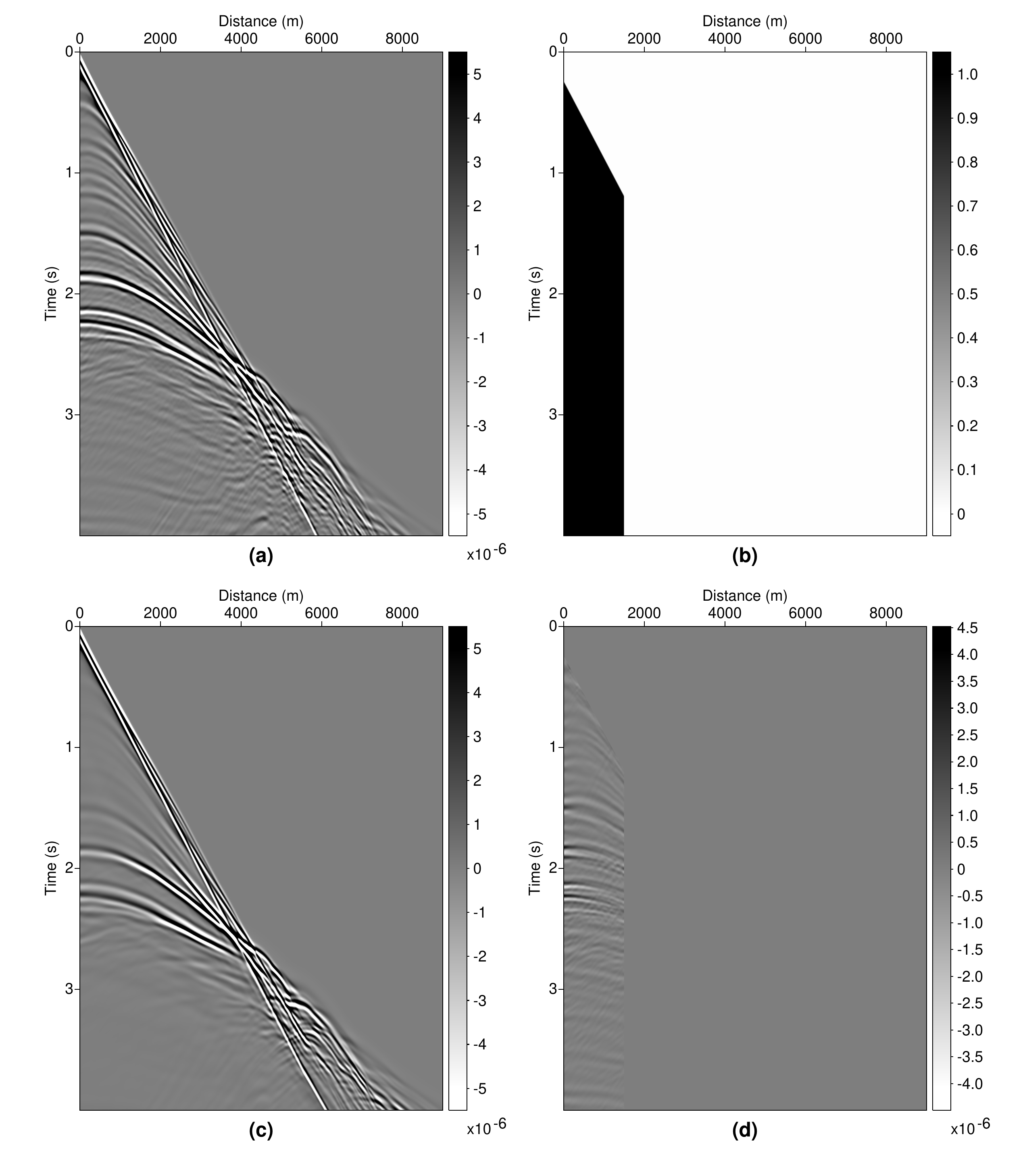}
  \caption{(a) the first shot  modelled from true Marmousi model; (b) muting mask; (c) simulated data from smooth Marmousi model; (d) the residual data as the input for linearized waveform inversion after muting direct arrivals and far-offset diving waves.}\label{fig:dat_0001}
\end{figure}

The observed seismograms (Figure~\ref{fig:dat_0001}a) are generated from the true model using a 12 Hz Ricker wavelet. The data comprise 4000 time steps with a sampling interval of 0.001 s. A muting mask (Figure~\ref{fig:dat_0001}b) is applied to isolate near-offset reflections from direct waves and long-offset refractions. Figure~\ref{fig:dat_0001}c shows the first shot gather simulated in the background medium. This synthetic data is subtracted from the observed data, and the result is muted before inversion. The final input for data-domain LSRTM is these muted data residuals, shown in Figure~\ref{fig:dat_0001}d.

\begin{figure}[htbp]
  \centering
  \includegraphics[width=\linewidth]{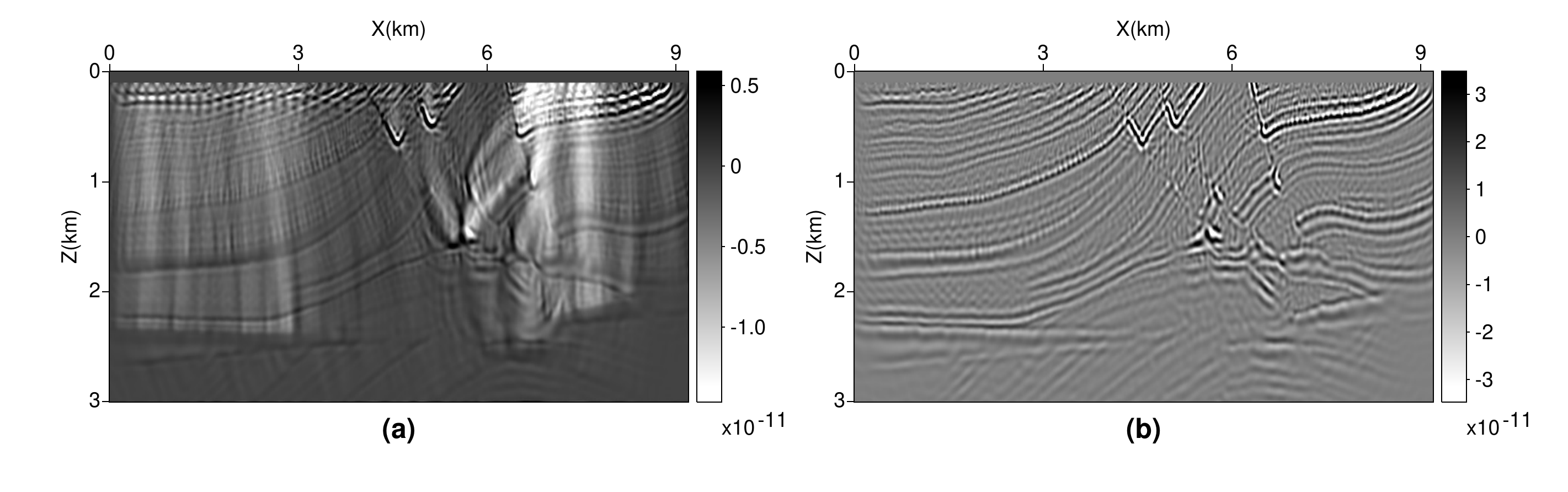}
  \caption{RTM images for logarithmic parameters: (a) wavespeed  $\ln V_p$ and (b) impedance $\ln I_p$}\label{fig:rtm}
\end{figure}

Migration of the reflection data produces the generalized RTM images for  $\ln V_p$ and $\ln I_p$ shown in Figure~\ref{fig:rtm}. It is important to note that these images are obtained directly from the cross-correlation of the source and receiver wavefields, without any post-processing filtering.
The $V_p$ image (Figure~\ref{fig:rtm}a) contains strong, low-wavenumber components along the wavepaths, whereas the impedance image (Figure~\ref{fig:rtm}b) is largely free of such low-frequency artifacts. This observation aligns with the finding of \citet{douma2010connection}, who demonstrated that the impedance kernel is mathematically equivalent to applying a Laplacian filter to a conventional RTM image.

\begin{figure}[htbp]
  \centering
  \includegraphics[width=0.65\linewidth]{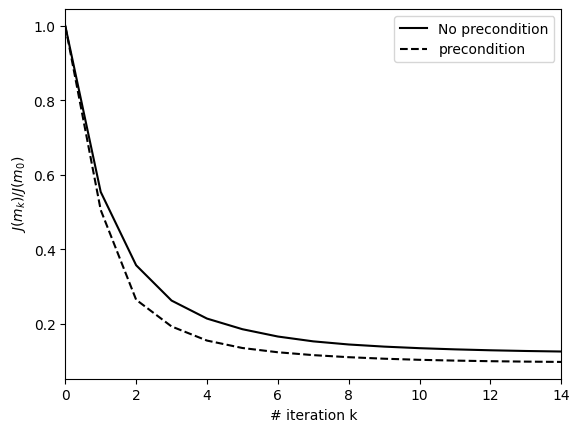}
  \caption{Normalized misfit of data-domain LSRTM with (PCGNR) and without (CGNR) preconditioning}\label{fig:misfit2d}
\end{figure}

  We performed data-domain LSRTM for 15 CGNR iterations, both with and without pseudo-Hessian preconditioning. The convergence history in Figure~\ref{fig:misfit2d} shows that the preconditioned CGNR (PCGNR) algorithm achieves a lower normalized misfit for the same number of iterations, indicating a superior convergence rate compared to the unpreconditioned approach.

\begin{figure}[htbp]
  \centering
  \includegraphics[width=\linewidth]{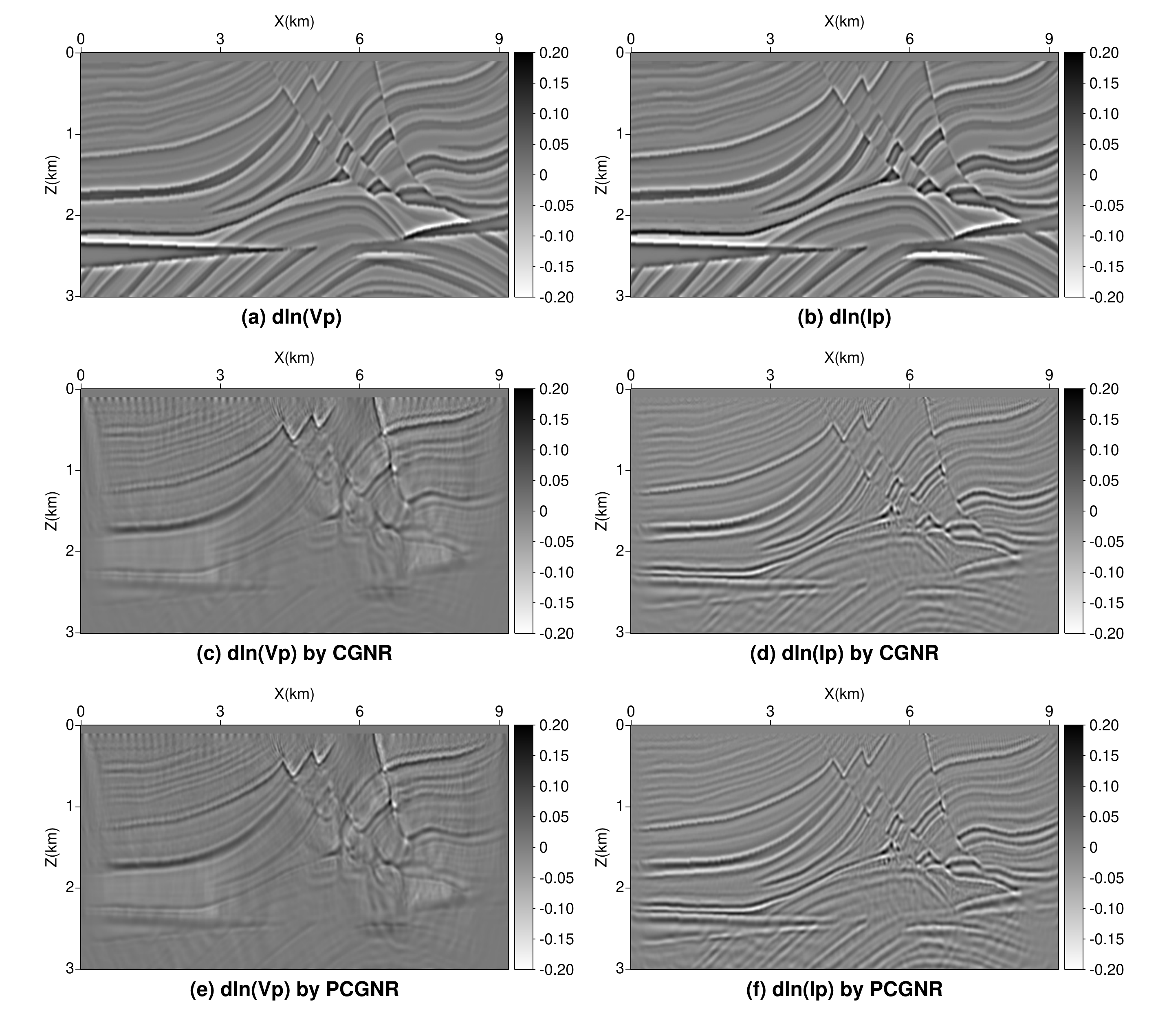}
  \caption{ (a, b) true $\delta \ln V_p$ and $\delta\ln I_p$;  (c, d)  $\delta \ln V_p$ and $\delta \ln I_p$ inverted by CGNR; (e, f)  $\delta \ln V_p$ and $\delta \ln I_p$ inverted by PCGNR.}\label{fig:dm2dinv}
\end{figure}

\begin{figure}[htbp]
  \centering
  \includegraphics[width=0.85\linewidth]{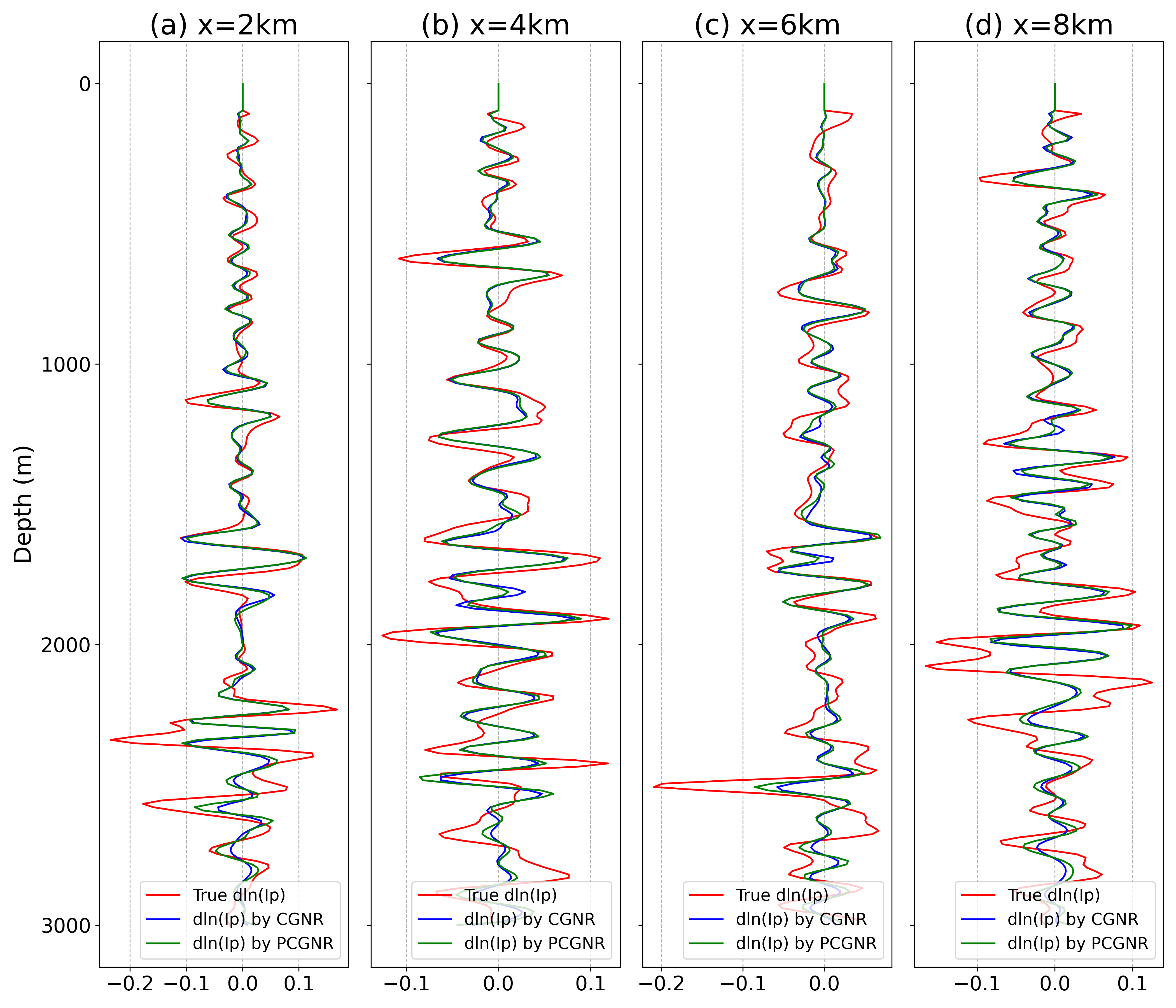}
  \caption{$\delta \ln I_p$ inverted in data domain at 4 different locations: (a) x=2 km; (b) x=4 km; (c) x=6 km; (d) x=8 km}\label{fig:welllog}
\end{figure}

  The final inverted model perturbations are displayed in Figure~\ref{fig:dm2dinv}. For reference, the true model perturbations for velocity and impedance are shown in Figure~\ref{fig:dm2dinv}a and ~\ref{fig:dm2dinv}b. A comparison reveals that the results from preconditioned LSRTM (Figure~\ref{fig:dm2dinv}e and ~\ref{fig:dm2dinv}f) exhibit sharper boundaries and higher amplitudes at depth than those from the unpreconditioned algorithm (Figure~\ref{fig:dm2dinv}c and ~\ref{fig:dm2dinv}d) after the same number of iterations.
  To further validate the results, we extracted the inverted $\delta\ln I_p$ at four locations ($x=2,4,6,8 km$) and compared them with the true perturbation in Figure~\ref{fig:welllog}. The PCGNR result aligns slightly closer with the true model than the CGNR result, though both show a high degree of consistency.

\subsection{3D Overthrust model}

\begin{figure}[htbp]
  \centering
  \includegraphics[width=0.9\linewidth]{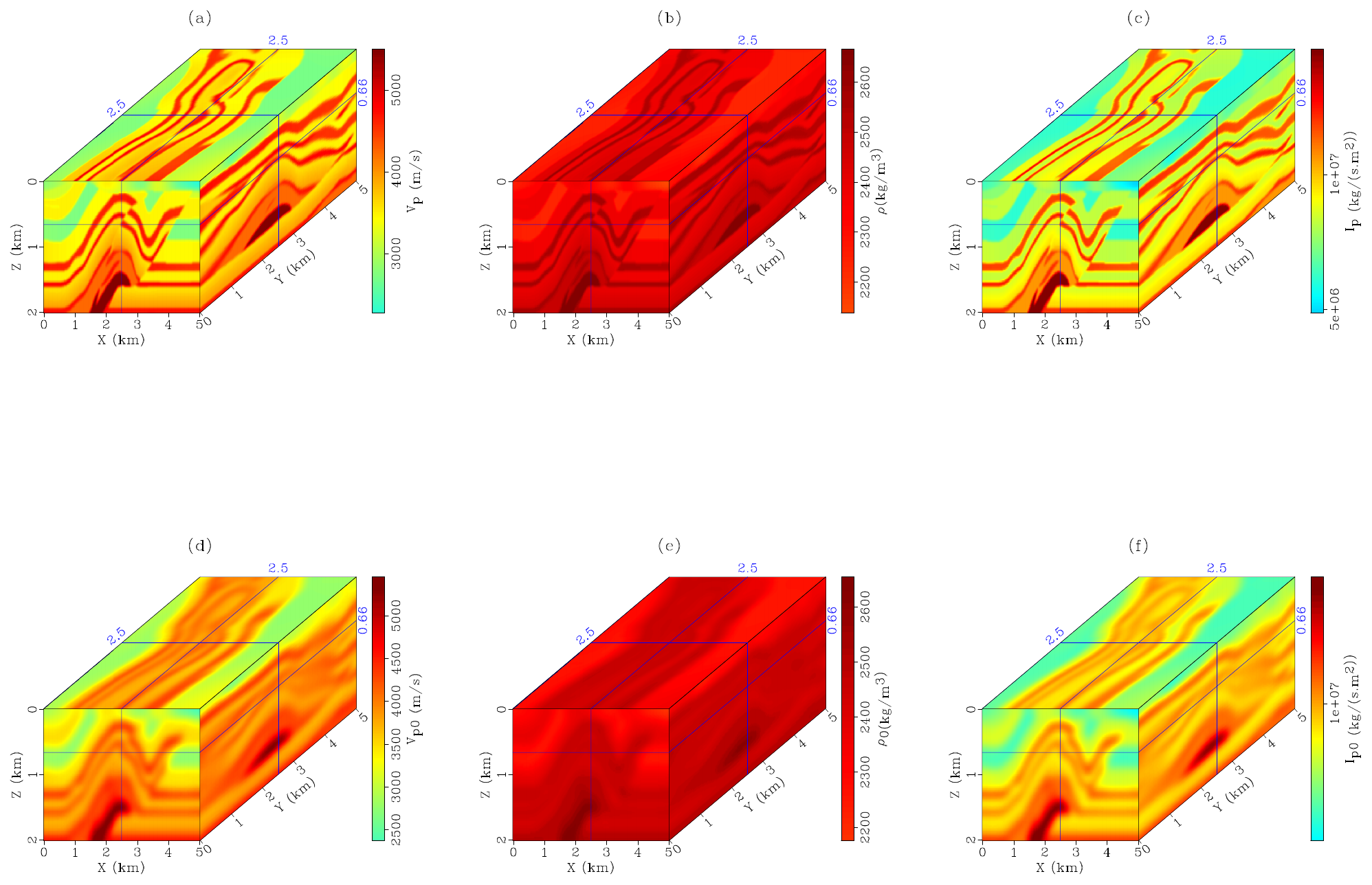}
  \caption{3D data-domain LSRTM models: The 1st row and the 2nd row are the true models and the background models corresponding to the parameters (a, d) velocity, (b, e) density and  (c,f) the actual impedance.}\label{fig:mod3d}
\end{figure}

We now test our multiparameter LSRTM on a subset of the 3D Overthrust model. The test volume has dimensions of $n_x\times n_y\times n_z=201\times 201\times 101$, with grid spacings of $\Delta x=\Delta y=25$ m and $\Delta z=20$ m.
Figure~\ref{fig:mod3d}a and ~\ref{fig:mod3d}b show the true velocity and density models, respectively. These are highly smoothed to create the reference models for LSRTM, shown in Figure~\ref{fig:mod3d}d and ~\ref{fig:mod3d}e. The corresponding true and initial impedance models, derived from these, are displayed in Figure~\ref{fig:mod3d}c and ~\ref{fig:mod3d}f.

\begin{figure}[htbp]
  \centering
  \includegraphics[width=0.65\linewidth]{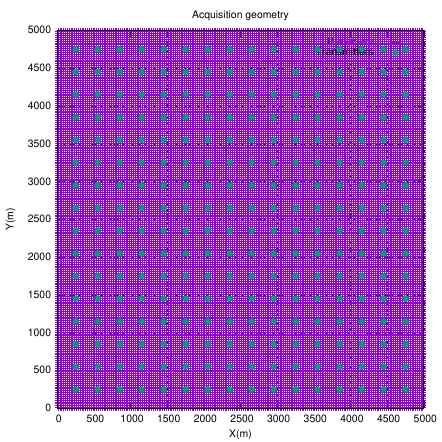}
  \caption{Source-receiver configuration for 3D data-domain LSRTM}\label{fig:acquisition3d}
\end{figure}

The acquisition geometry is extensive, comprising 256 shots. Each shot is recorded by 40,000 receivers, all evenly distributed in the horizontal plane; the source-receiver layout is illustrated in Figure~\ref{fig:acquisition3d}. Sources and receivers are placed at depths of 5 m and 10 m, respectively. Numerical modeling uses a 12 Hz Ricker wavelet and runs for 1500 time steps with a sampling interval of  $\Delta t=0.0018$ s.

\begin{figure}[htbp]
  \centering
  \includegraphics[width=\linewidth]{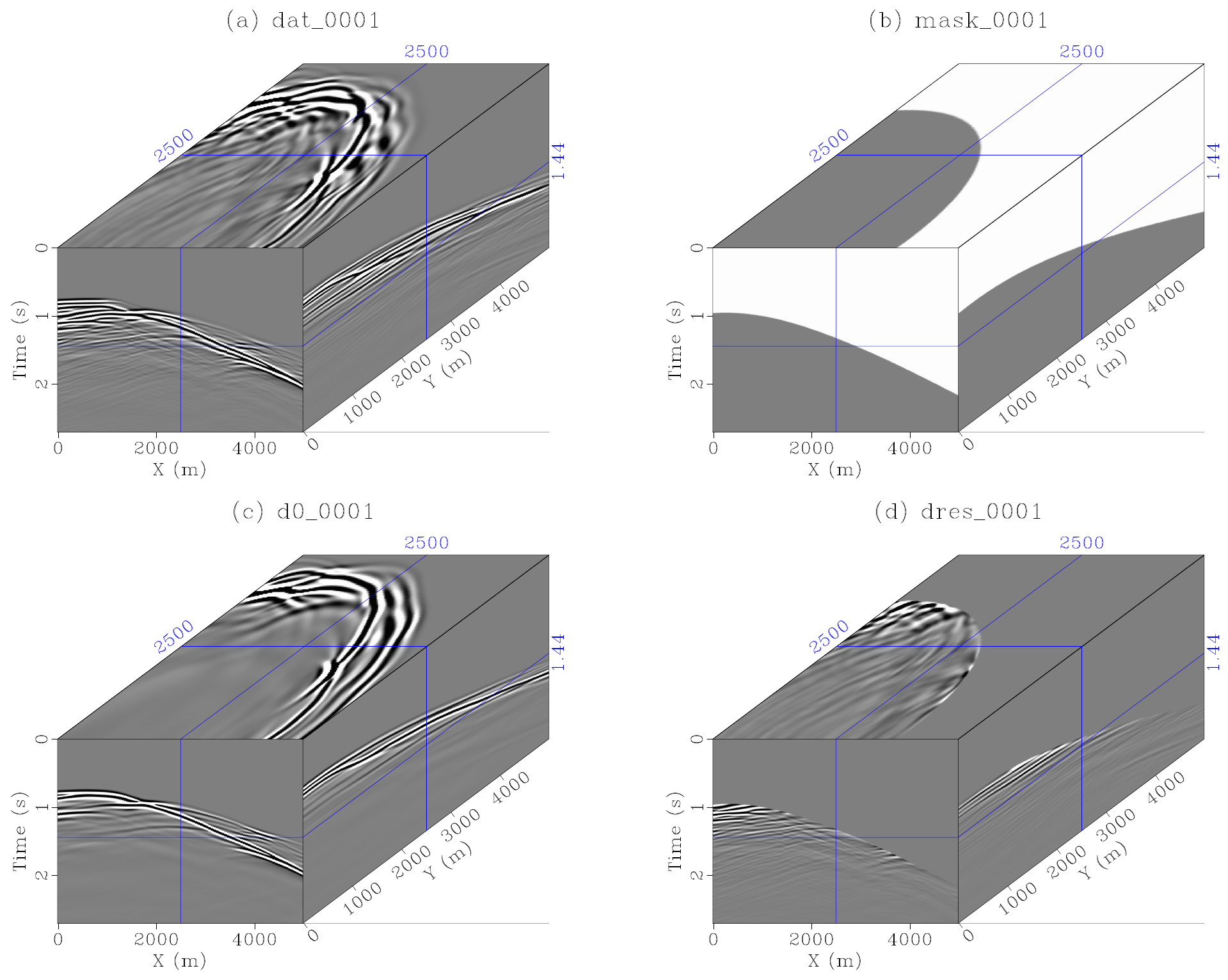}
  \caption{(a) The 1st shot (complete data modelled from true model); (b) muting mask; (c) modelled data from background medium; (d) data residual $\delta d$ used for linearized inversion}\label{fig:dat3d}
\end{figure}

\begin{figure}[htbp]
  \centering
  \includegraphics[width=\linewidth]{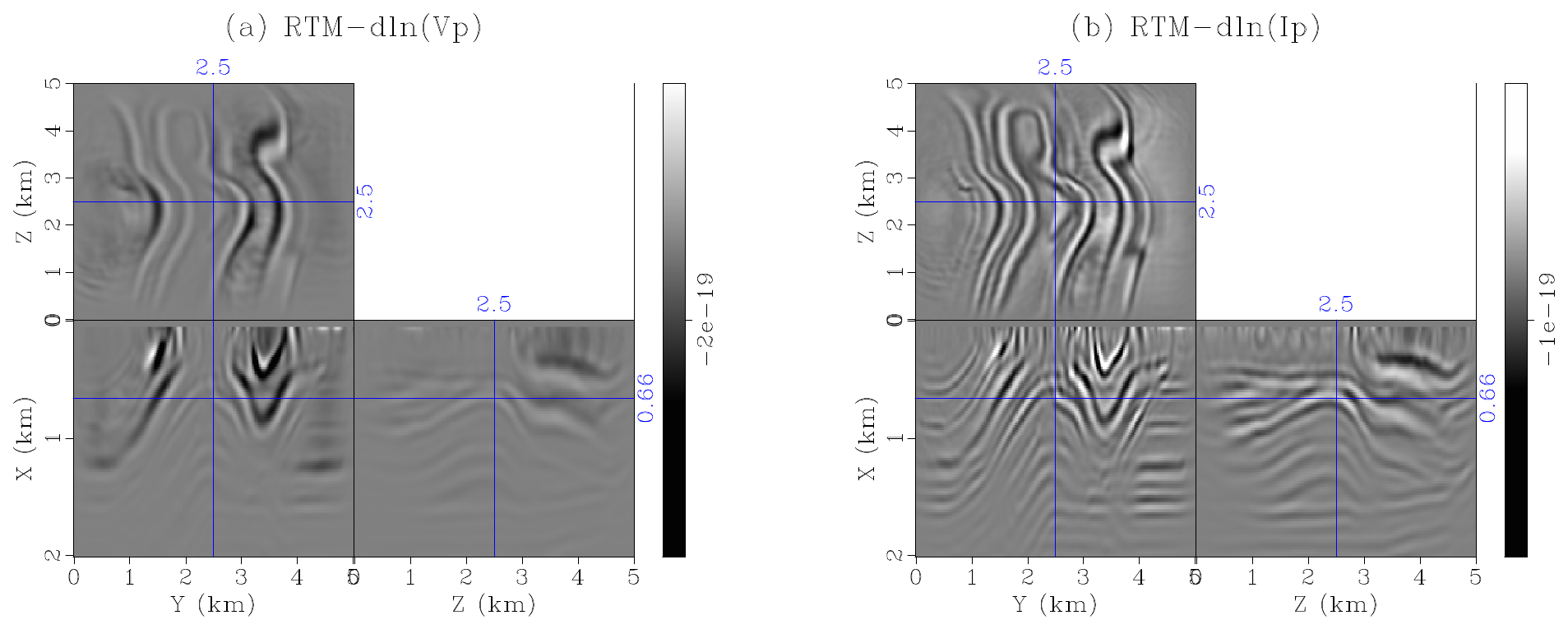}
  \caption{RTM images for parameters (a) $\ln V_p$ and (b) $\ln I_p$}\label{fig:rtm3d}
\end{figure}

\begin{figure}[htbp]
  \centering
  \includegraphics[width=\linewidth]{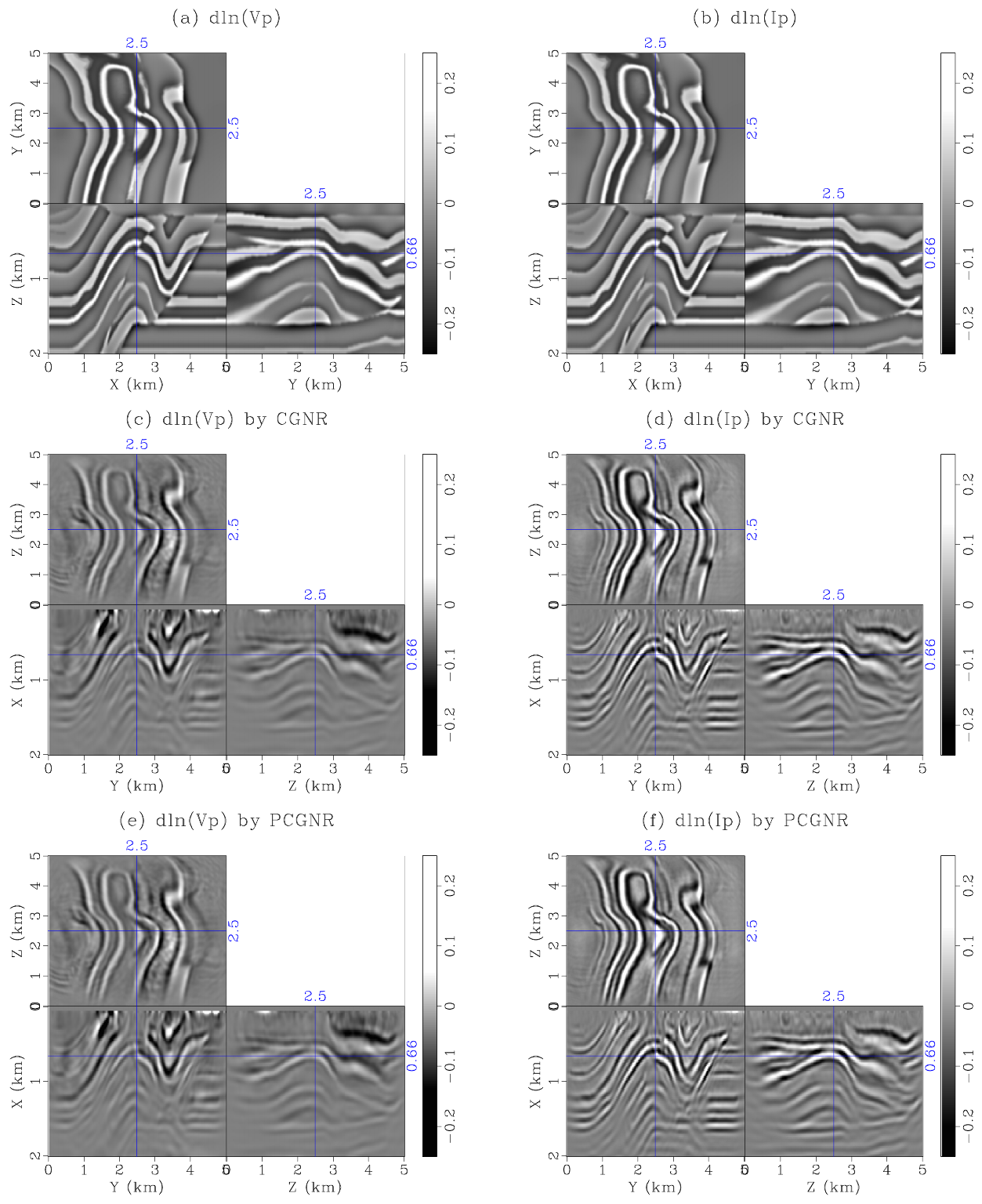}
  \caption{(a,b) True model perturbations for velocity and impedance; (c,d) model perturbations inverted by unpreconditioned LSRTM; (e,f) model perturbations inverted by preconditioned LSRTM}\label{fig:dm3d}
\end{figure}

\begin{figure}[htbp]
  \centering
  \includegraphics[width=0.65\linewidth]{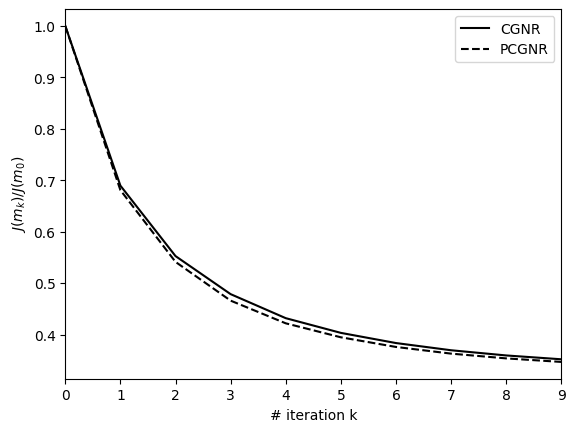}
  \caption{Normalized misfit of data-domain LSRTM with (PCGNR) and without (CGNR) preconditioning for 3D Overthrust model}\label{fig:misfit3d}
\end{figure}

Figure~\ref{fig:dat3d} illustrates the 3D seismic data, showing (a) the first shot gather, (b) the applied muting mask, (c) the synthetic data modeled in the reference medium, and (d) the final muted data residuals used for linearized inversion.
The initial RTM images for the velocity and impedance parameters are presented in Figure~\ref{fig:rtm3d}, which exhibit relatively low resolution. After 10 iterations of LSRTM, the image quality is dramatically improved, successfully recovering the amplitude of the model perturbations. As shown in Figure~\ref{fig:dm3d}, the preconditioned LSRTM result provides a marginally better reflectivity image, particularly in the deeper section of the model, compared to the unpreconditioned approach. This is consistent with the convergence history in Figure~\ref{fig:misfit3d}, where the preconditioned CGNR (PCGNR) algorithm achieves a lower normalized misfit than the standard CGNR for the same number of iterations. These findings align well with the results from the previous 2D test.

\section{Distinctions of the methods}

\subsection{Data-domain versus image-domain}

\begin{figure}[htbp]
  \centering
  \includegraphics[width=\linewidth]{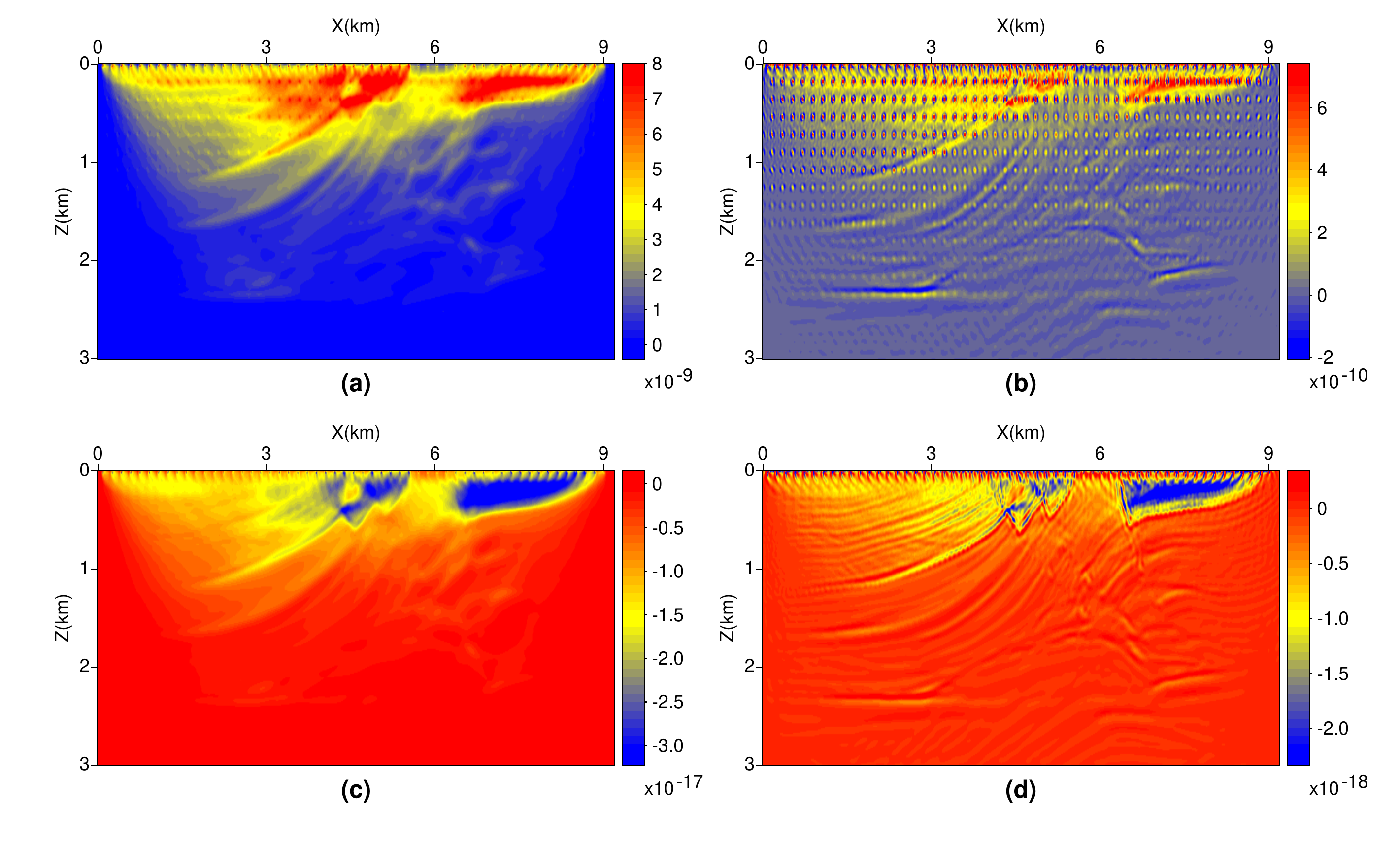}
  \caption{(a, b) PSFs computed by evenly (every 15 points) distributing a number of scatter points in lateral and vertical directions and (c, d) re-migrated image $m''$ associated with different parameters: (a,c) velocity and (b,d) impedance}\label{fig:m2par}
\end{figure}

\begin{figure}[htbp]
  \centering
  \includegraphics[width=\linewidth]{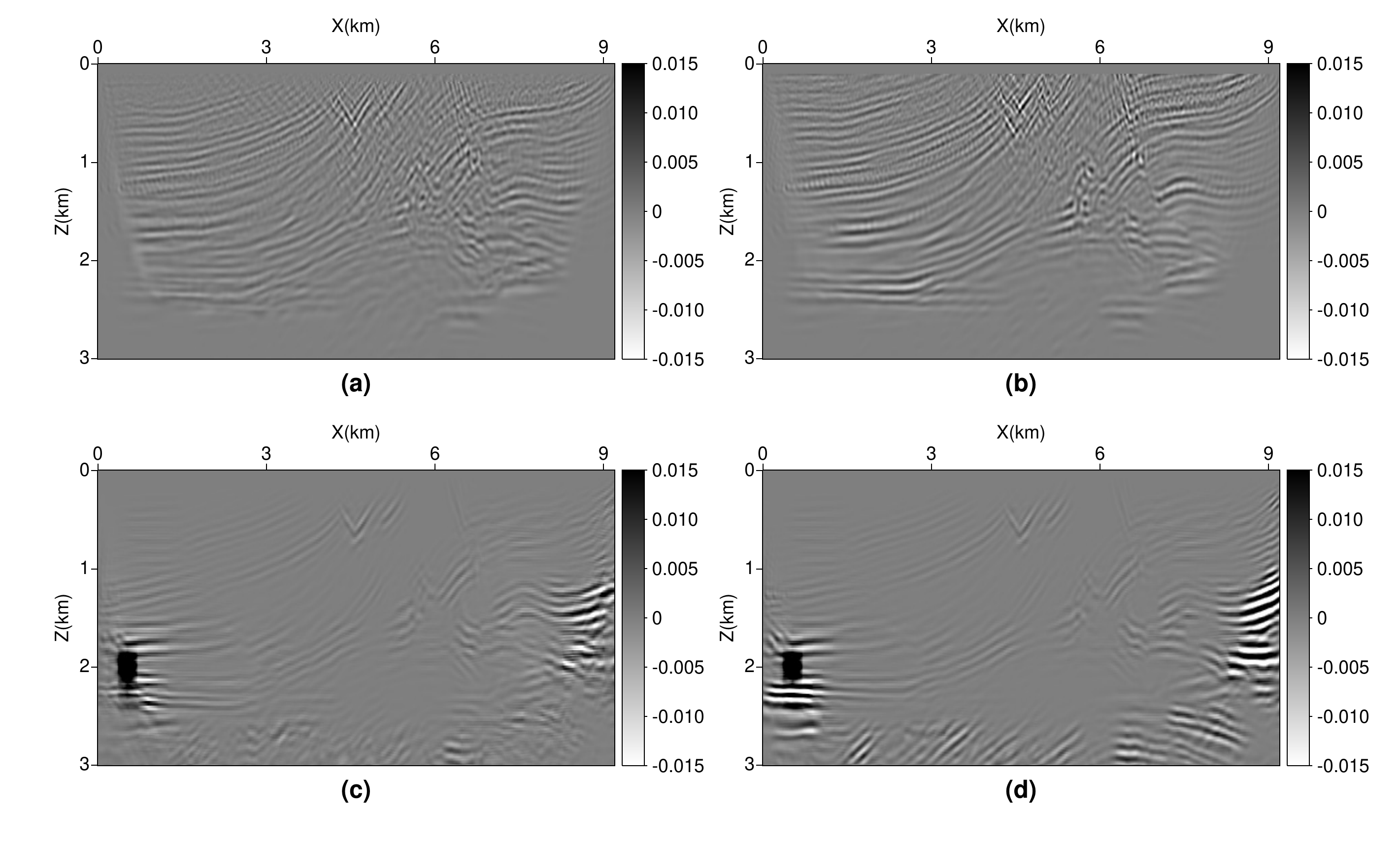}
  \caption{Model perturbation computed by multiparameter image-domain LSRTM using (a,b) matrix-based PSF Hessian and (c,d) FFT-based Wiener filtering for two parameters (a, c) $V_p$ and (b, d) $I_p$ (the large values at the dark areas in c and d are due to clipping)}\label{fig:decon2par}
\end{figure}

Based on the 2D Marmousi model presented previously, we computed the PSFs for velocity and impedance parameters in Figure~\ref{fig:m2par}a and \ref{fig:m2par}b. The PSFs for velocity shown in Figure~\ref{fig:m2par}a bear strong tomographic modes with long propagation path, while the PSFs for impedance shown in Figure~\ref{fig:m2par}b have much better focusing effect and can be spatially separated. Using these sparsely sampled PSFs, we are able to compute Hessian vector product by interpolation of missing columns on the fly. After 100 CGNR iterations, we reach a normalized misfit value approximately 0.05. Though the misfit reaches a small enough level (cf. Figure~\ref{fig:psfconv}), the model perturbations after linear inversion are quite unsatisfactory, as shown in Figure~\ref{fig:decon2par}.
For deblurring filter approach, the remigrated RTM images corresponding to velocity and impedance are shown in Figure~\ref{fig:m2par}c and \ref{fig:m2par}d. Combined with the RTM image before remigration, we then estimate nonstationary Wiener filter using FFT to do deblurring. The reflectivity images obtained after filtering does not provide any useful information about the structure of the subsurface, as can be seen in Figure~\ref{fig:decon2par}c and \ref{fig:decon2par}d.

\subsection{Monoparameter versus multiparameter inversion}

\begin{figure}[htbp]
  \centering
  \includegraphics[width=\linewidth]{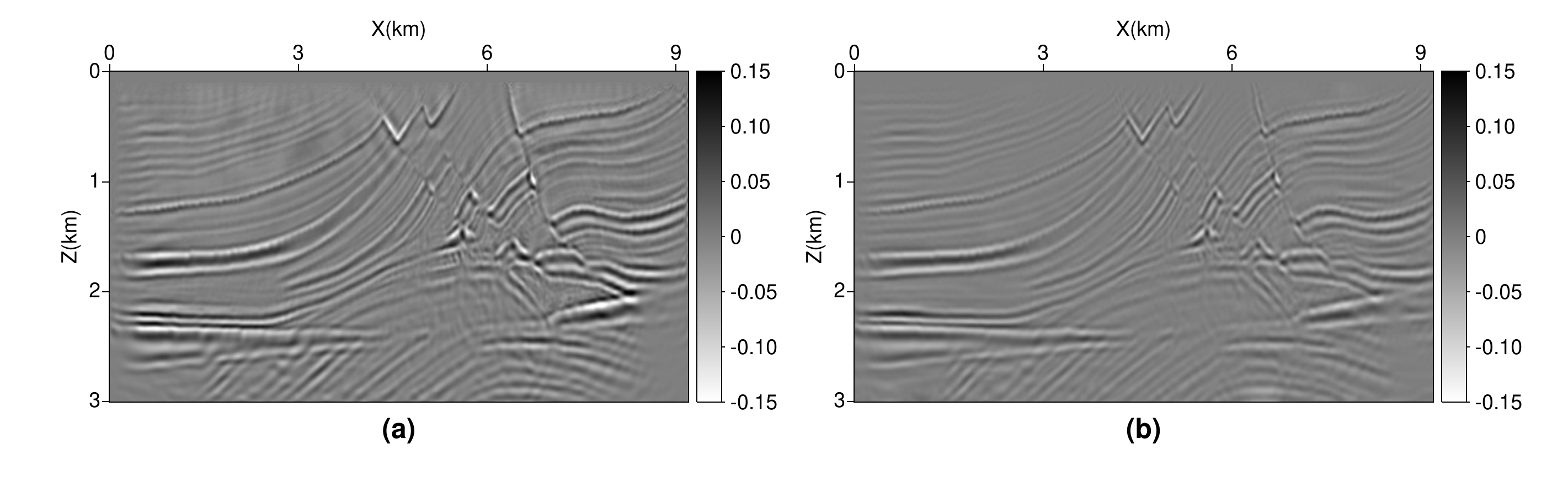}
  \caption{Model perturbation computed by monoparameter image-domain LSRTM using (a) matrix-based PSF Hessian and (b) FFT-based Wiener filtering for impedance parameter only}\label{fig:decon1par}
\end{figure}

\begin{figure}[htbp]
  \centering
  \includegraphics[width=0.65\linewidth]{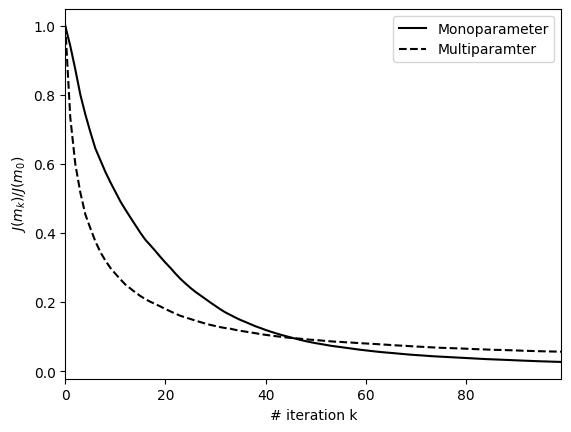}
  \caption{The convergence history of matrix-based migration deconvolution with monoparameter and multiparamter inversion}\label{fig:psfconv}
\end{figure}

\begin{figure}[htbp]
  \centering
  \includegraphics[width=0.85\linewidth]{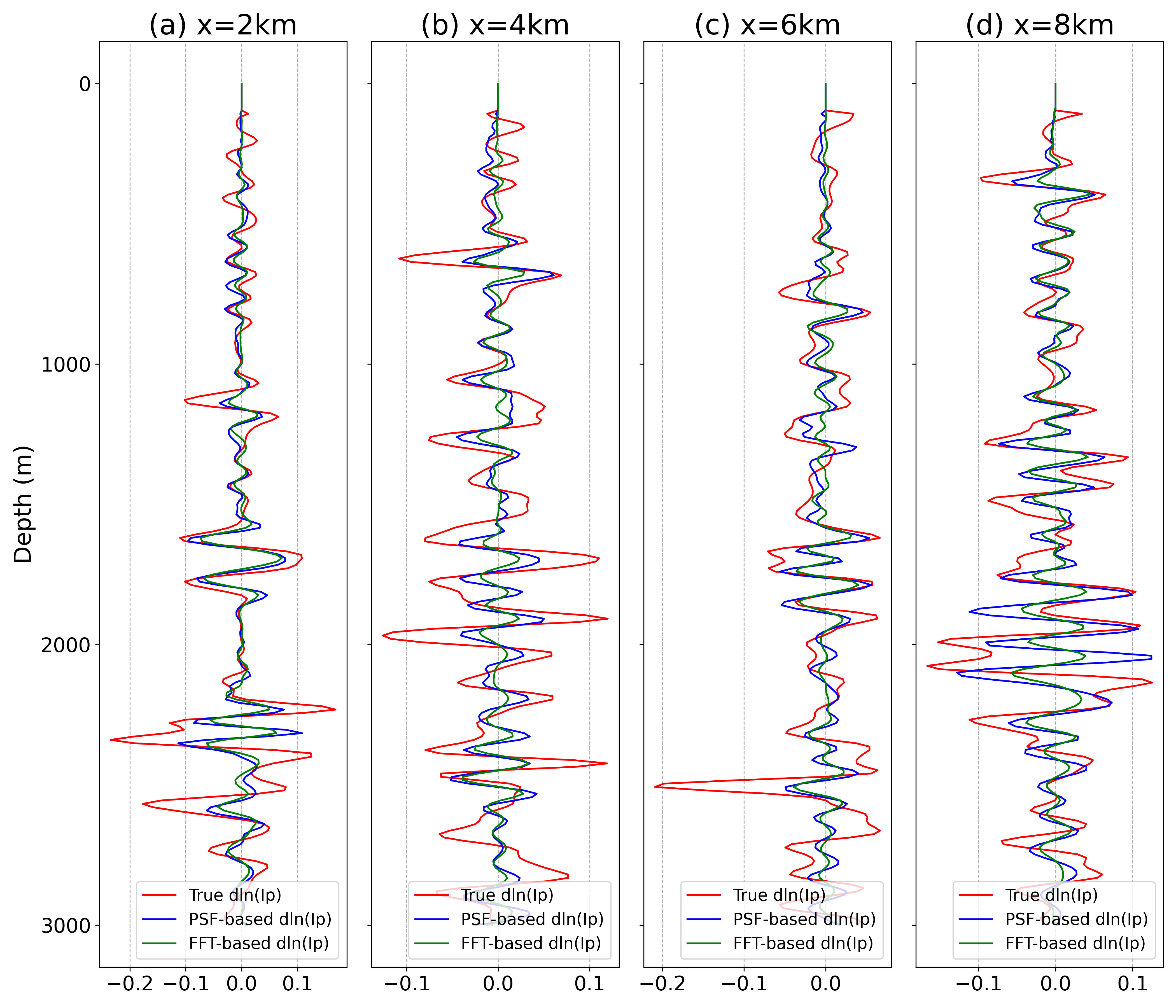}
  \caption{$\delta \ln I_p$ inverted in monoparameter image domain at 4 different locations: (a) x=2 km; (b) x=4 km; (c) x=6 km; (d) x=8 km}\label{fig:welllog2}
\end{figure}

The above results reveal that multiparameter image-domain LSRTM using interpolated PSF Hessian and nonstationary Wiener filter cannot return the reflectivity images of the same quality as data-domain approach, due to strong spatial correlation of the PSFs for parameters and possibly insufficient sampling of the Hessian. 
We now switch to monoparameter mode to concentrate on impedance perturbation only. Figure~\ref{fig:decon1par}a and \ref{fig:decon1par}b show the reflectivity images of the impedance retrieved by image-domain LSRTM using matrix-based PSF Hessian and nonstationary Wiener filter, respectively.  Clearly, the images inverted in monoparameter mode are slightly poorer than the result obtained by data-domain approach, but significantly better than the images computed in multiparameter image-domain mode. This is again confirmed from the well logs  extracted at the same spatial locations in comparison between Figure~\ref{fig:welllog} and Figure~\ref{fig:welllog2}. The normalized misfit for monoparameter migration deconvolution finally reaches a satisfactory level, as can be seen in Figure~\ref{fig:psfconv}. 

In fact, the multiparameter inversion is mathematically different from monoparameter inversion. Recall the inverse of partitioned matrix 
\begin{equation}
  \begin{bmatrix}
    H_{11} & H_{12}\\
    H_{21} & H_{22}
  \end{bmatrix}^{-1}=\begin{bmatrix}
  (H_{11}-H_{12}H_{22}^{-1}H_{21})^{-1} & -(H_{11}-H_{12}H_{22}^{-1}H_{21})^{-1}H_{12}H_{22}^{-1}\\
  -H_{22}^{-1}H_{21}(H_{11}-H_{12}H_{22}^{-1}H_{21})^{-1} & H_{22}^{-1}+H_{22}^{-1}H_{21}(H_{11}-H_{12}H_{22}^{-1}H_{21})^{-1}H_{12}H_{22}^{-1}
  \end{bmatrix}.
\end{equation}
The estimated image in two-parameter mode is thus
\begin{equation}
  \begin{split}
    \begin{bmatrix}
      \delta m_1\\
      \delta m_2
    \end{bmatrix}=&\begin{bmatrix}
      H_{11} & H_{12}\\
      H_{21} & H_{22}
    \end{bmatrix}^{-1}\begin{bmatrix}
      m_{rtm,1}\\
      m_{rtm,2}
    \end{bmatrix}
    =\begin{bmatrix}
    (H_{11}-H_{12}H_{22}^{-1}H_{21})^{-1} m_{rtm,1}-(H_{11}-H_{12}H_{22}^{-1}H_{21})^{-1}H_{12}H_{22}^{-1} m_{rtm,2}\\
    H_{22}^{-1} m_{rtm,2}-H_{22}^{-1}H_{21}(H_{11}-H_{12}H_{22}^{-1}H_{21})^{-1}(m_{rtm,1}-H_{12}H_{22}^{-1}m_{rtm,2})
    \end{bmatrix}
  \end{split}
\end{equation}
The monoparameter inversion with Hessian $H_{22}$ for the RTM image $m_{rtm,2}$ is
\begin{equation}
  \delta m_2 = H_{22}^{-1} m_{rtm,2}.
\end{equation}
From the above we clearly see that the coupling footprint from parameter 1 neglected in monoparameter inversion corresponds to
\begin{equation}
  -H_{22}^{-1}H_{21}(H_{11}-H_{12}H_{22}^{-1}H_{21})^{-1}(m_{rtm,1}-H_{12}H_{22}^{-1}m_{rtm,2}).
\end{equation}

\subsection{Compromise over computational efficiency, memory overhead and imaging accuracy}

In terms of computational cost, image-domain least-squares reverse-time migration (LSRTM) requires only approximately twice the cost of a single RTM. The primary expense in the iterative PSF-based method lies in the matrix-vector products at each iteration, while the cost of the subsequent Wiener filtering is negligible. This represents a significant advantage over data-domain LSRTM, which requires repetitive wave modeling for a massive number of independent sources.

A central challenge for image-domain LSRTM is the requirement of the Hessian matrix, which is prohibitively large to store for large-scale problems. The PSF approach approximates the Hessian as a sparse, banded matrix. Storage issues can be further alleviated by on-the-fly interpolation, which relies on the assumption that PSFs are well-separated with minimal dispersed energy. However, this assumption only holds if the migration velocity model is sufficiently smooth and lacks high contrasts.
In practice, migration models for real media inevitably contain contrasts and cannot be perfectly smooth. Consequently, the precise construction of the Hessian via spatial interpolation is somewhat contradictory to the fundamental assumption required for linearized inversion to be applicable.

Improving the quality of multiparameter inversion via PSF-based migration deconvolution would likely require storing and accessing more elements of the Hessian, ideally computed via an analytical approach. Unfortunately, an explicit expression for the multiparameter Hessian is absent from the literature, despite the monoparameter Hessian for velocity perturbation derived by \citet{Plessix_2004_FDF} being widely used \citep{valenciano2006target,zhang2024angle}.

Even for monoparameter LSRTM, Figure~\ref{fig:m2par}b indicates that constructing the Hessian requires estimating spatially varying PSFs (i.e., the density of point scatterers should differ from the top to the bottom of the model). A possible remedy, suggested by \citet{alger2024point}, uses a high-rank Hessian approximation, but this would significantly complicate the numerical implementation and is left for future work. Furthermore, the local translation invariance of PSFs breaks down in regions with strong velocity contrasts. As a result, the nonstationary Wiener filtering approach is expected to yield a poor estimation of model perturbations in such areas.

\section{Application to Viking Graben dataset}

We applied our method to the Viking Graben Line 12 dataset from the North Sea Basin \citep{keys1998data}. In preparation for linearized imaging, we compensated for spherical divergence by scaling the data by $\sqrt{t}$. Due to computational constraints and the bijective shot-CPU mapping in SMIwiz, we subsampled the dataset from its original 1001 shots (25 m spacing) to 85 shots with 300 m spacing. Each shot comprises 120 channels, recording a 6 s time series at a 4 ms sampling rate.

\begin{figure}[htbp]
  \centering
  \includegraphics[width=\linewidth]{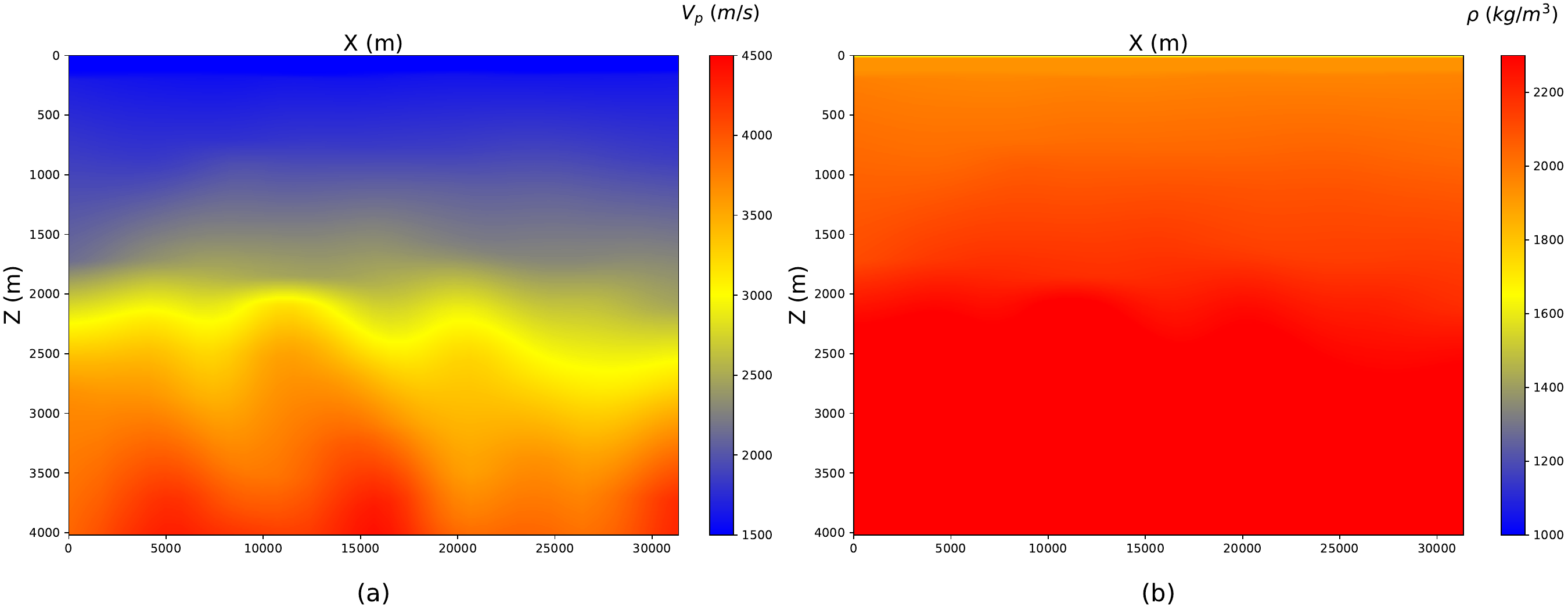}
  \caption{The migration velocity and density for 2D Viking Graben Line 12 dataset}\label{fig:vprho}
\end{figure}

\begin{figure}[htbp]
  \centering
  \includegraphics[width=0.9\linewidth]{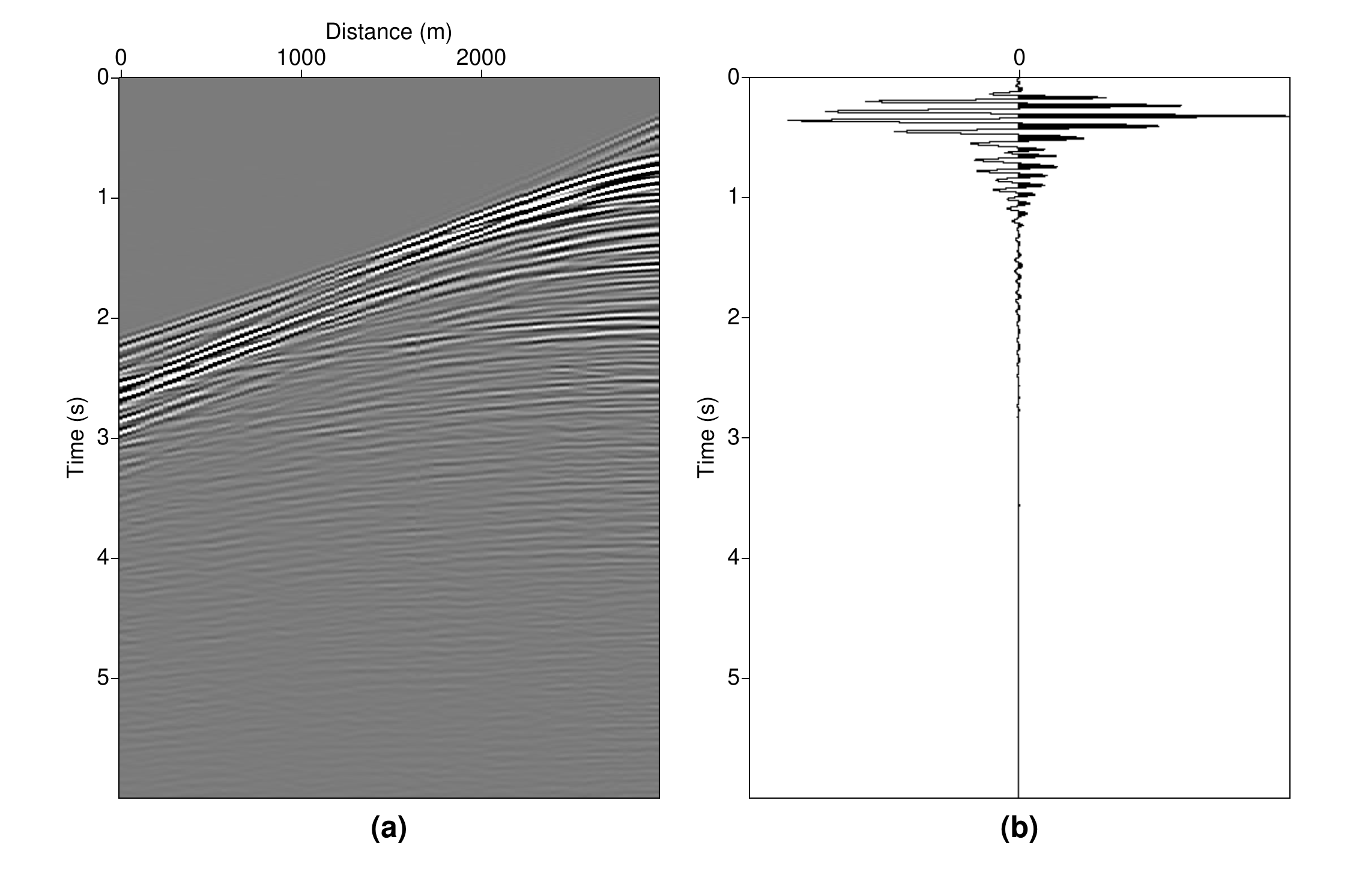}
  \caption{(a) The first shot after lowpass filtering and (b) the estimated source wavelet}\label{fig:datwlt}
\end{figure}

Migration velocity analysis, performed at 39 locations, yielded a smooth velocity model with a minimum of 1500 m/s (in water) and a maximum of approximately 4410 m/s. This model was interpolated onto a grid with $\Delta x=\Delta z=20$ m (Figure~\ref{fig:vprho}a). We then derived a density model (Figure~\ref{fig:vprho}b) from the velocity model using Gardner's law \citep{gardner1974formation}.
Although the data contain frequencies up to 60 Hz, the dispersion relation limited the usable bandwidth to a maximum of 15 Hz. Similar to \citet{Yang_2018_EAGE}, we applied a minimum-phase lowpass filter and estimated a source wavelet from the filtered data using the method of \citet{Pratt_1999_SWIb}. The first shot gather and the estimated wavelet are shown in Figure~\ref{fig:datwlt}.

\begin{figure}[htbp]
  \centering
  \includegraphics[width=\linewidth]{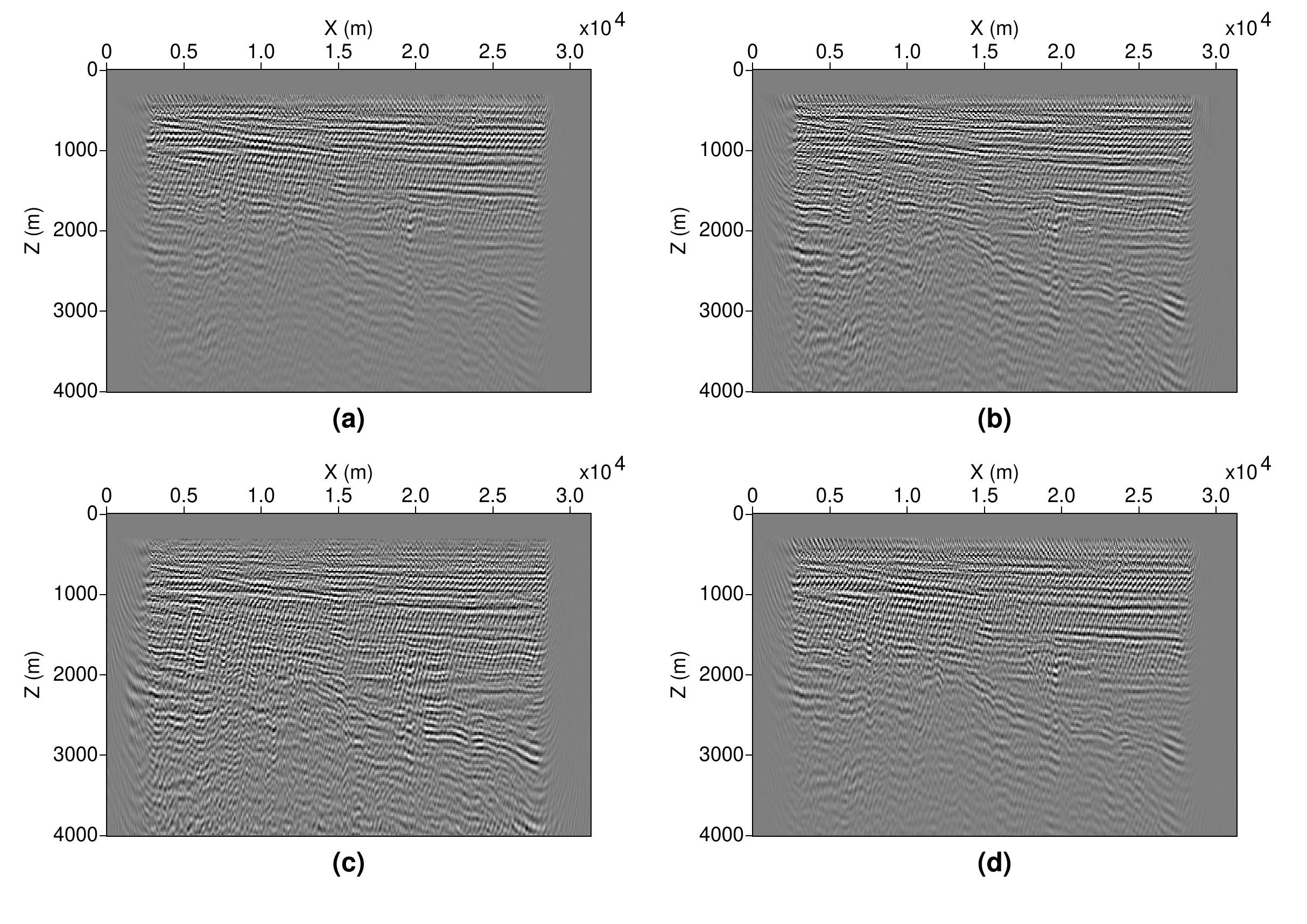}\\
  \caption{Viking Graben imaging result computed from (a) RTM, (b) data-domain LSRTM, and image-domain LSRTM using (c) PSF and (d) nonstationary deblurring filter}\label{fig:vikingimage}
\end{figure}

Using this wavelet, we performed two-parameter reverse-time migration (RTM) and least-squares RTM (LSRTM). As shown in Figure~\ref{fig:vikingimage}, the impedance image from data-domain LSRTM (Figure~\ref{fig:vikingimage}b) is significantly better than that from RTM (Figure~\ref{fig:vikingimage}a), with highly improved illumination at depth.
We also computed impedance images via monoparameter image-domain LSRTM using point-spread functions (PSF) and a deblurring filter (Figure~\ref{fig:vikingimage}c and \ref{fig:vikingimage}d). The quality of the monoparameter image-domain results is also comparable to the data-domain counterpart; they appear somewhat distorted due to the rigid window size and limited aperture. While this could potentially be improved by tuning parameters such as window size and overlap, such tuning can be subjective.

It is important to note that a large portion of frequency content embedded in the data up to 60 Hz has not been exploited, while less than 1/10 shots were used  in the above imaging experiment. Since our initial velocity and density models are quite inaccurate, we anticipate dramatic improvement of imaging resolution by refining the migration velocity model using advanced FWI techniques \citep{cui2024low} before proceeding with RTM and LSRTM. Moving to higher frequencies for imaging at refined grid spacing using all seismic shots is expected to improve the continuity of the seismic events in the resulting image.

\section{Conclusion}

We have implemented acoustic multiparameter LSRTM in both the image and data domains, using a velocity-impedance parameterization and log scaling based on the first-order velocity-pressure formulation of the wave equation. This work includes, to our knowledge, the first reported instance of 3D multiparameter data-domain LSRTM with this parameterization. The Viking Graben example demonstrates the validity of our implementation within the  framework of SMIwiz open-source software.

A comparison between the two domains reveals their distinctive characteristics. Data-domain LSRTM generally provides a quantitative estimation of model perturbations with higher resolution than its image-domain counterpart, albeit at a significantly higher computational cost. While image-domain LSRTM faces considerable challenges in multiparameter linearized inversion, it remains viable for inverting a single impedance parameter, as the corresponding point-spread functions (PSFs) remain locally focused.

We observed that pseudo-Hessian preconditioning provided only a limited acceleration in convergence rate under this parameterization, despite its widespread promotion in other contexts. The results presented here could be substantially improved by employing better initial models and more refined parameter tuning. Furthermore, an extension from acoustic to elastic physics, incorporating seismic attenuation and anisotropy, is conceptually straightforward in SMIwiz.

\section*{Acknowledgements}

This research is financially supported by National Natural Science Foundation of China (Grant No. 42274156). Pengliang Yang appreciates the discussions with Jean Virieux, Yujin Liu and  Wei Zhang on image-domain LSRTM. We acknowledge the valuable help from Wei Zhou on handling the Viking Graben dataset, and the  positive feedback from the audiences at EAGE 2025 in Toulouse when presenting this work publically \citep{yang2025comparison}.

\section*{Computer Code Availability}

The code has been released publically via github repository \url{https://github.com/yangpl/SMIwiz}. The examples in this paper are reproducible under the directories \verb|run_lsrtm2d| and \verb|run_lsrtm3d|. The example of Viking Graben Line 12 data set, processed with Seismic Unix commands, is reproducible under the directory \verb|run_Viking|.

\appendix
\section{The adjoint of Born modelling}\label{sec:mig}

The Lagrangian functional translates the linear minimization with scattering equation constraint into an unconstrained optimization
\begin{equation}
  \mathcal{L}(\delta m, \delta u, \lambda) =\frac{1}{2}\|R\delta u- \delta d\|^2
  +\langle\lambda, A(m_0)\delta u_0 + \delta m\cdot \partial_m A(m_0) u_0 \rangle_{X\times[0,T]},
\end{equation}
where we have introduced the Lagrange multiplier $\lambda(\mathbf{x},t)$ and equipped with the inner product defined as follows:
\begin{equation}
  \langle h, g\rangle_{X\times[0,T]} = \int_X  \int_0^T  h(\mathbf{x},t)^\mathrm{H} g(\mathbf{x},t) \mathrm{d}\mathbf{x}\mathrm{d}t, \quad \mathbf{x}\in X, \; t\in [0, T].
\end{equation}
The gradient of the Lagrangian with respect to $\delta m$ is then
\begin{equation}\label{eq:partialLpartialmp}
  \frac{\partial \mathcal{L}}{\partial \delta m}
  =\frac{\partial \mathcal{L}}{\partial \lambda}\frac{\partial \lambda}{\partial \delta m}
  +\frac{\partial \mathcal{L}}{\partial \delta u}\frac{\partial \delta u}{\partial \delta m}
  +\langle \lambda, \frac{\partial(\delta m\cdot \partial_m A(m_0))}{\partial \delta m} u_0\rangle_{[0,T]}.
\end{equation}
Zeroing the partial derivatives of the Lagrangian with respect to $\lambda$ ($\partial\mathcal{L}/\partial \lambda=0$) gives exactly the forward scattering equation \eqref{eq:scattereq}. Zeroing the partial derivatives of the Lagrangian with respect to $\delta u$ ($\partial\mathcal{L}/\partial \delta u=0$) gives the adjoint equation
\begin{equation}\label{eq:adjeq}
  A^\mathrm{H} \lambda =R^\mathrm{H} (\delta d- R\delta u), 
\end{equation}
which implies that the adjoint field $\lambda$ must be simulated using the residual of the reflection data as the virtual sources.
Let us split adjoint field and data residual into particle velocity  and pressure, say, $\lambda=[\lambda_\mathbf{v}, \lambda_p]^\mathrm{H}$ and $ R^\mathrm{H} (\delta d-R\delta u)=[\Delta d_\mathbf{v},\Delta d_p]^\mathrm{H}$. 
Since $(\partial_t)^\mathrm{H} = -\partial_t$, $(\nabla)^\mathrm{H} = -\nabla\cdot$ and  $(\kappa\nabla\cdot)^\mathrm{H}=-\nabla \kappa$, the adjoint equation in equation \eqref{eq:adjeq} is
\begin{equation}\label{eq:adjnotused}
  \begin{bmatrix}
    -\rho\partial_t & -\nabla\kappa\\
    -\nabla\cdot & -\partial_t
  \end{bmatrix}\begin{bmatrix}
    \lambda_\mathbf{v}\\
    \lambda_p
  \end{bmatrix} = \begin{bmatrix}
    \Delta d_\mathbf{v}\\
    \Delta d_p
  \end{bmatrix}.
\end{equation}
Define $\bar{\mathbf{v}}=\lambda_\mathbf{v}$ and $\bar{p}=\kappa\lambda_p$. Equation~\eqref{eq:adjnotused} translates into
\begin{equation}
  \begin{bmatrix}
    \rho\partial_t & \nabla\\
    \kappa\nabla\cdot & \partial_t
  \end{bmatrix}\begin{bmatrix}
    \bar{\mathbf{v}}\\
    \bar{p}
  \end{bmatrix} = \begin{bmatrix}
    -\Delta d_\mathbf{v}\\
    -\kappa\Delta d_p
  \end{bmatrix},
\end{equation}
where the wave operator becomes the same as the forward operator thanks to a compliance formulation \citep{Yang_2016_SFM,Yang_2018_TRN}. This allows the same forward modelling code to be reused for adjoint simulation, as long as the adjoint source is properly scaled during the backward injection of the weighted data residual. 

At the saddle point, both the forward scattering equation and the adjoint equation are satisfied.
In terms of \eqref{eq:partialLpartialmp}, the gradient of the objective with respect to $\delta m$ is then
\begin{equation}\label{eq:gradlsm}
  \frac{\partial J(\delta m)}{\partial \delta m}= \langle \lambda, \partial_m A(m_0) u_0\rangle_{[0,T]}
  =\int_0^T \mathrm{d}t \lambda^\mathrm{H}(\mathbf{x},t) \frac{\partial A(m_0)}{\partial m} u_0(\mathbf{x},t),
\end{equation}
where the stacking over sources and receivers are omitted henceforth. Let us point out that the adjoint equation must be simulated backwards in time, opposite to the forward simulation. The computation of the gradient in \eqref{eq:gradlsm} requires acessing both forward background field $u(\mathbf{x},t)$ and the adjoint field $\lambda(\mathbf{x},t)$ at all time steps. This creates a key computational challenge. Therefore, a number of computing schemes have been proposed, such as lossy data compression \citep{Sun_2013_TWO,Boehm_2015_WCA}, wavefield reconstruction via stored boundaries \citep{Yang_2016_WRB} and optimal checkpointing strategy \citep{Symes_2007_RTM,Yang_2016_CAR}.
In view of the relation \eqref{eq:ln}, the gradient of the misfit with respect to density and bulk modulus after log scaling are
\begin{subequations}
  \begin{align}
    \frac{\partial J(\delta m)}{\partial \delta \ln\rho}=&\rho\int_0^T \mathrm{d}t \begin{bmatrix}
      \lambda_\mathbf{v}^\mathrm{H} &\lambda_p\end{bmatrix}\begin{bmatrix}
        \partial_t & 0\\
        0 & 0
      \end{bmatrix}\begin{bmatrix}
        \mathbf{v}\\
        p
      \end{bmatrix}
      =\rho\int_0^T \lambda_\mathbf{v}^\mathrm{H}\partial_t \mathbf{v} \mathrm{d}t
      =\rho\int_0^T \bar{\mathbf{v}}^\mathrm{H}\partial_t \mathbf{v} \mathrm{d}t,\label{eq:gradrho}\\
      \frac{\partial J(\delta m)}{\partial \delta\ln\kappa}=&\kappa\int_0^T  \mathrm{d}t \begin{bmatrix}
        \lambda_\mathbf{v}^\mathrm{H} &\lambda_p\end{bmatrix}\begin{bmatrix}
          0 & 0\\
          \nabla\cdot & 0
        \end{bmatrix}\begin{bmatrix}
          \mathbf{v}\\
          p
        \end{bmatrix}
        =\kappa\int_0^T \lambda_p\nabla\cdot \mathbf{v}\mathrm{d}t
        =\int_0^T \bar{p}\nabla\cdot \mathbf{v}\mathrm{d}t.\label{eq:gradkappa}
  \end{align}
\end{subequations}

\section{Pseudo-Hessian preconditioning}\label{sec:pseudohessian}

To improve the convergence rate of the linearized inversion, pseudo-Hessian preconditioner proposed in \citet{Shin_2001_IAP} approximates the diagonal elements of the Hessian up to a scaling factor
\begin{equation}
  \tilde{H}_{m,m}=\int_0^T \mathrm{d}t \left(\frac{\partial A}{\partial m}u_0\right)^\mathrm{H} \left(\frac{\partial A}{\partial m}u_0\right)
\end{equation}
by replacing the receiver field with the source field based on the assumption of zero-offset geometry. In view of equation \eqref{eq:ln}, we arrive at
\begin{subequations}
  \begin{align}
    \tilde{H}_{\ln\rho,\ln\rho} =& \int_0^T \mathrm{d}t \left(\frac{\partial A}{\partial\ln\rho}u_0\right)^\mathrm{H}\left(\frac{\partial A}{\partial\ln\rho}u_0\right)
    =\rho^2\int_0^T \mathrm{d}t (\partial_t \mathbf{v})\cdot(\partial_t \mathbf{v}),\\
    \tilde{H}_{\ln\kappa,\ln\kappa} =& \int_0^T \mathrm{d}t \left(\frac{\partial A}{\partial\ln\kappa}u_0\right)^\mathrm{H}\left(\frac{\partial A}{\partial\ln\kappa}u_0\right)
    =\kappa^2\int_0^T \mathrm{d}t (\nabla\cdot \mathbf{v})^2.
  \end{align}
\end{subequations}
Using the relation in equation \eqref{eq:chainrule2}, the diagonal elements of the pseudo-Hessian under $\ln V_p-\ln I_p$ parametrization are
\begin{equation}
  \tilde{H}_{\ln V_p,\ln V_p}=\tilde{H}_{\ln I_p,\ln I_p}=\tilde{H}_{\ln\rho,\ln\rho} + \tilde{H}_{\ln\kappa,\ln\kappa}.
\end{equation}
Since constructing pseudo-Hessian only requires the incident background field without any dependence on the change of model perturbation, this preconditioner can be computed only once and used through all iterations of linear inversion.

\section{Migration deconvolution using PSF}\label{sec:psf}

The key to migration deconvolution is to access the Hessian matrix $H$.  A simple idea  is to sample it by point scatter (or reflectivity spike), corresponding to Dirac delta function. Using such a point scatter as the input perturbation parameter, the output after one migration and one demigration, will be nothing else than a specific column of the Hessian matrix, which is often referred to PSF \citep{lecomte2008resolution,fletcher2016least}. In theory, the full Hessian could be formed by stacking all PSFs from every grid point, which is computationally intractable. As a practical solution, we distribute a number of distant point scatters all at a time.  Application of a migration and a demigration gives an image vector which is then the superposition of many PSFs.

\begin{figure}[htbp]
  \centering
  \includegraphics[width=0.8\linewidth]{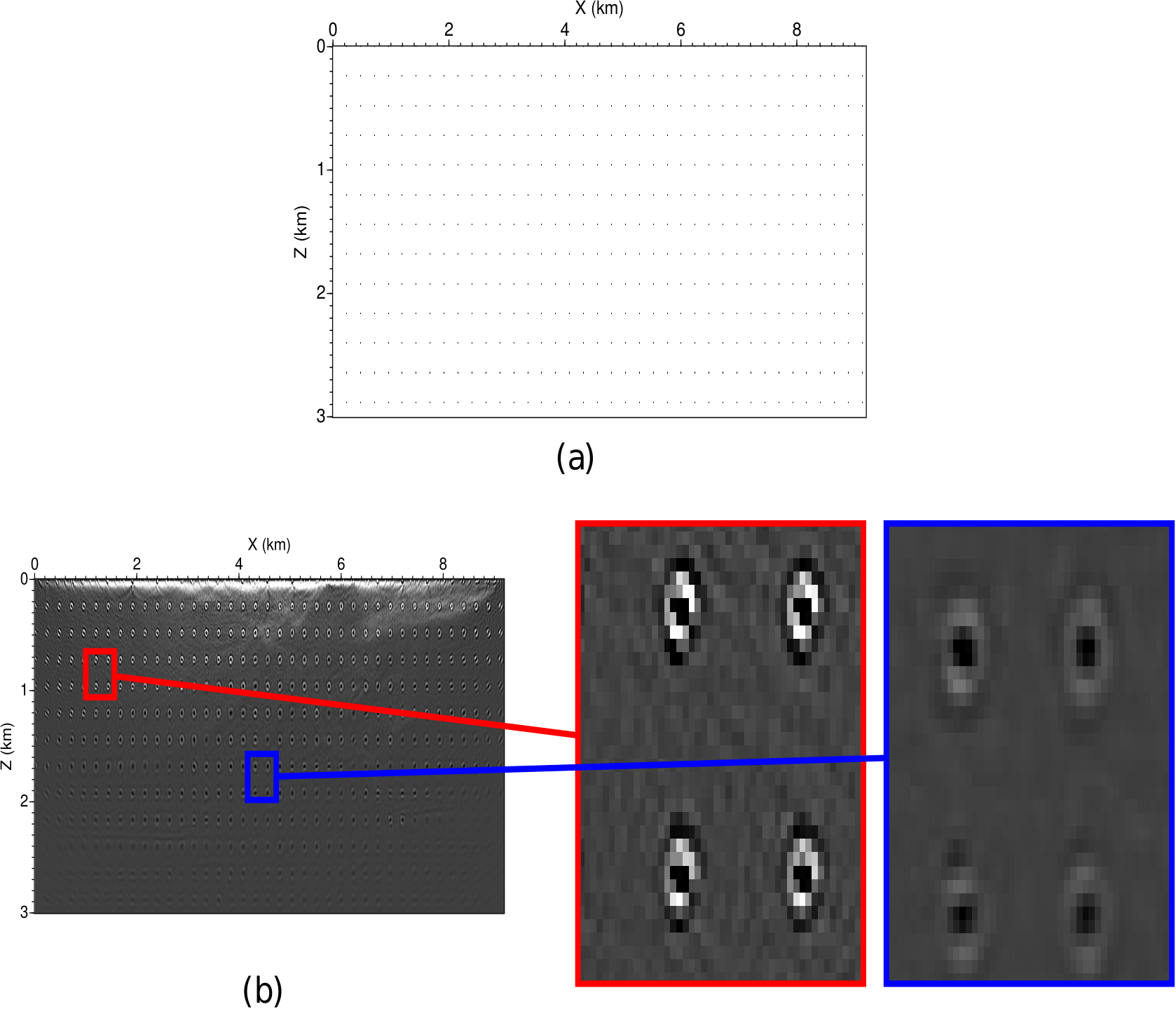}
  \caption{(a) point scatters evenly distributed one after every 20 points in lateral and vertical directions; (b) the resulting PSF image showing focused energy surrounding the scattering point.}\label{fig:apppsf}
\end{figure}

Figure~\ref{fig:apppsf}a shows a number of scatter points for sampling Hessian in a Marmousi model of size $151\times 461$. The resulting PSF image after one migration and one demigration is plotted in Figure~\ref{fig:apppsf}b. Each scatter point generates to a localized image, highly focused with very limited spatial support.
This validates the argument that Hessian is approximately a scaled, bandlimited identity operator \citep{gray1997true,Symes_2008_MVA}.
It implies that we can safely neglect the elements of PSF outside a small window of every localized image point, representing this specific column of Hessian using very few non-zero values. Though stored in the same image compactly, the focused image point at different locations will correspond to another column of the Hessian matrix. This means that the PSF image sparsely samples few columns of Hessian.

\begin{figure}[htbp]
  \centering
  \includegraphics[width=0.8\linewidth]{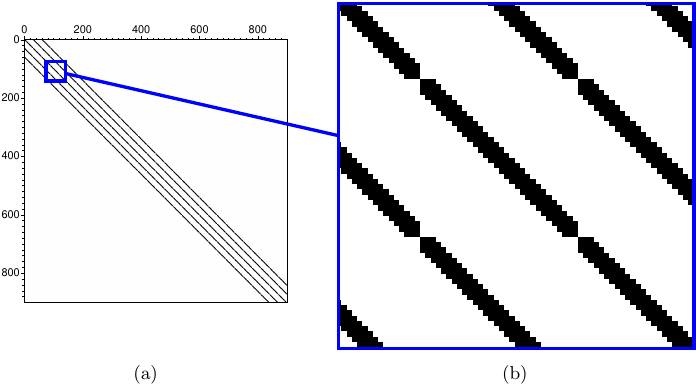}
  \caption{(a) The sparsity pattern of interpolated PSF Hessian associated with a model of size $30\times 30$ with scatters distributed every 5 points in x and y directions (5 band groups in total with each group including 5 bands); (b) A zoom-in display of the non-zero bands.}\label{fig:bandedmatrix}
\end{figure}

Figure~\ref{fig:apppsf}b highlights that the image points next to each other share high geometrical similarity and amplitudes, indicating that the translation invariance is locally valid in space. This suggests that the other columns of Hessian which correspond to a location in between and have not yet been sampled by distributing a scatter point can be approximately estimated though trilinear interpolation, based on the sampled image points at 8 corners of a cube surrounding it \citep{osorio2021migration}.
For any image point $P_{i,j,k}$ with 8 surrounding corner nodes $(i_{0/1}, j_{0/1}, k_{0/1})$ ($i_0\leq i < i_1$, $j_0\leq j< j_1$, $k_0\leq k<k_1$), we have
\begin{equation}\label{eq:trilinear}
  \begin{split}
 P_{i,j,k} =& (1-w_1)(1-w_2)(1-w_3) P_{i_0,j_0,k_0} + w_1(1-w_2)(1-w_3) P_{i_1,j_0,k_0} \\
  +& (1-w_1)w_2(1-w_3) P_{i_0,j_1,k_0} + w_1w_2(1-w_3) P_{i_1,j_1,k_0}\\
+&(1-w_1)(1-w_2)w_3 P_{i_0,j_0,k_1} + w_1(1-w_2)w_3 P_{i_1,j_0,k_1} \\
+& (1-w_1)w_2w_3 P_{i_0,j_1,k_1} + w_1w_2w_3 P_{i_1,j_1,k_1},
\end{split}
\end{equation}
where the interpolating weights are 
\begin{equation}
  w_1 = \frac{x_i-x_{i_0}}{x_{i_1}-x_{i_0}},
  w_2 = \frac{y_j-y_{j_0}}{y_{j_1}-y_{j_0}},
  w_3 = \frac{z_k-z_{k_0}}{z_{k_1}-z_{k_0}}.
\end{equation}
The trilinear interpolation degenerates to bilinear interpolation in 2D. It allows us to access all columns of Hessian matrix, whose sampled elements form a sparse banded matrix, see an illustration in Figure~\ref{fig:bandedmatrix}.

The PSF Hessian $H$ constructed in such a way gives a parsimonious representation of the complete Hessian $H$. Unfortunately, the partial sampling and interpolation result in the loss of SPD properties in PSF Hessian. Therefore, we resort to CGNR algorithm to solve the normal equation of problem  \eqref{eq:migdeconobj} $H^\mathrm{H} H \delta m=H^\mathrm{H} m_{rtm}$,
where both PSF Hessian $H$ and its adjoint $H^\mathrm{H}$ must be used for constructing matrix-vector multiplication. Indeed, only the action of the matrix to an input vector is required in CGNR algorithm. Thanks to the geometrical similarity shown in Figure~\ref{fig:apppsf}, the interpolation could therefore be done on the fly to compute matrix vector product, avoiding significant consumption of computer memory while maintaining efficiency.
The columns of the Hessian matrix computed via trilinear interpolation are linear combination of some others, indicating that the PSF Hessian is not invertible. The minimization of the problem in \eqref{eq:migdeconobj} gives a minimum energy solution in least-squares sense.

\section{Nonstationary Wiener filtering}\label{sec:fft}

Assume the Hessian matrix $H$ is locally invariant in space. This means the element $H_{ij}$ depends only on the relative shift $i-j$, resulting in a circulant Toeplitz structure of $H$. It can then be diagonalized by Fourier transform. The eigenvalue decomposition of Hessian  can then be expressed $H= F^\mathrm{H}\Lambda F$ \cite[chapter 5.2, corollary 5.16]{Vogel_2002_TVR},
where the Fourier matrix $F$ formed by the eigenvectors is unitary, i.e. $F^\mathrm{H} F=I$ ($F$ and $F^\mathrm{H}=F^{-1}$ are linear operators corresponding to forward and inverse Fourier transforms). Equation \eqref{eq:normal} can then be converted to
\begin{equation}\label{eq:sol}
  \delta m= F^\mathrm{H}\Lambda^{-1} Fm_{rtm},
\end{equation}
where $\Lambda^{-1}$ has to be found to estimate $\delta m$.

Migration of the Born modelled data  from a given image $m'$ yields an image
$m''=L^\mathrm{H}[m_0] L[m_0] m'= H m'$.
Applying Fourier transform on both sides yields $  \Lambda^{-1} Fm''= F m'$, or equivalently $F^{-1}[\Lambda^{-1}] * m''=  m'$, which reveals that the two images $m'$ and $m''$ are indeed connected via a spatial convolutional filter $F[\Lambda^{-1}]$.
Because the matrix $\Lambda$ (hence $\Lambda^{-1}$) is diagonal, it corresponds to the pointwise multiplication in wavenumber domain. One can then formulate a regularized least-squares minimization problem to estimate $\Lambda^{-1}$
\begin{equation}
  \min_{\Lambda^{-1}} \|Fm'-\Lambda^{-1} F m''\|^2 +\epsilon \|\Lambda^{-1}\|^2=
  \min_{\Lambda_i^{-1}}\sum_i\|[Fm']_i-\Lambda_i^{-1} [Fm'']_i\|^2 + \epsilon \|\Lambda_i^{-1}\|^2,
\end{equation}
where  $\epsilon$ is the Tikhonov regularization parameter;  $\Lambda_i$ is the $i$th diagonal element/eigenvalue of $\Lambda$. The solution is then given by a Wiener filtering procedure in Fourier domain $\Lambda_i^{-1}=  [\overline{Fm''}]_i [Fm']_i /([\overline{Fm''}]_i[Fm'']_i +\epsilon)$, where the overbar indicates complex conjugate and $\epsilon$ serves as a stabilization factor to avoid division by zero. 
The above relation provides us a recipe to find $\Lambda^{-1}$ according to $m'$ and $m''$, and in turn yields the solution of the problem computed via equation \eqref{eq:sol}. In practice, we simply take RTM image $m_{rtm}$ as the input referece image $m'$. 
The final reflectivity image $\delta m$ is found by filtering through equation \eqref{eq:sol} without any iterations.

The spirt of the above method is nothing else than performing a scaling of the RTM image in the frequency domain (or phase space), which dates back to \citet{claerbout1994spectral} and was then further refined by  \citet{rickett2003illumination,guitton2004amplitude,Symes_2008_ALI}. Let us remark that the concept of deblurring filter is not new at all, see for typical examples \citep{aoki2009fast,dai2011least}. Despite the wide use of Fourier transform in migration deconvolution \citep{lecomte2008resolution}, this is the first time that the assumption of circulant Toeplitz structure of Hessian approximation is recognized behind such a method, to the best of our knowledge.

The translation invariance of the PSF is only locally valid in space. Obviously, this assumption is questionable when applied to the whole Hessian matrix. In practice,  migration deconvolution is carried out using local blocks of Hessian, leading to the estimation of nonstationary filters to incoporate the spatial variations   \citep{guitton2004amplitude}. To avoid the checkboard effect of nonstationary Wiener filtering, a proper weighting strategy for overlapping domain is essential $\delta m= \sum_i w_i  \delta m_i/\sum_i w_i$, where $w_i$ is the valence for local image pixel $\delta m_i$.

\bibliographystyle{cas-model2-names}
\newcommand{\SortNoop}[1]{}

\end{document}